\documentclass[12pt,a4paper,oneside]{article}
\usepackage{graphicx,amssymb,amsmath,color}
\usepackage[linkcolor={blue},citecolor={red},colorlinks=true]{hyperref}
\usepackage{enumerate}
\usepackage[margin=2cm]{geometry}
\usepackage{slashed}
\usepackage{multirow}

\def\beq{\begin{equation}}
\def\eeq{\end{equation}}
\def\bea{\begin{eqnarray}}
\def\eea{\end{eqnarray}}

\def\bit{\begin{itemize}}
\def\eit{\end{itemize}}

\def\l{\left}
\def\r{\right}

\def\baa{\begin{array}}
\def\eaa{\end{array}}

\def\d{\partial}

\def\simgt{\mathrel{\lower2.5pt\vbox{\lineskip=0pt\baselineskip=0pt
           \hbox{$>$}\hbox{$\sim$}}}}
\def\simlt{\mathrel{\lower2.5pt\vbox{\lineskip=0pt\baselineskip=0pt
           \hbox{$<$}\hbox{$\sim$}}}}

\def\bfc{\begin{figure}\begin{center}}
\def\efc{\end{center}\end{figure}}
\def\nn{\nonumber\\}

\begin{document}

\begin{flushright}
\hspace{3cm} 
SISSA  03/2020/FISI
\end{flushright}
\vspace{.6cm}
\begin{center}

\hspace{-0.4cm}{\Large \bf 
Phase transitions in perturbative walking dynamics}\\[0.5cm]

\vspace{1cm}{Aleksandr Azatov$^{a,b,c,1}$,  and Miguel Vanvlasselaer$^{a,b,c,2}$}
\\[7mm]
 {\it \small

$^a$ SISSA International School for Advanced Studies, Via Bonomea 265, 34136, Trieste, Italy\\[0.15cm]
$^b$ INFN - Sezione di Trieste, Via Bonomea 265, 34136, Trieste, Italy\\[0.1cm]
$^c$ IFPU, Institute for Fundamental Physics of the Universe, Via Beirut 2, 34014 Trieste, Italy\\[0.1cm]
 }

\end{center}

\bigskip \bigskip \bigskip

\centerline{\bf Abstract} 
\begin{quote}
In this paper, we investigate the dynamics of the confinement-deconfinement 
phase transition in a toy model where the walking dynamics is realized 
perturbatively. We study the properties of the phase transition focusing on the 
possible cosmological signatures it can provide. Interestingly the model is well 
under perturbative control only when the mass of the lightest field -  the dilaton/scalon is much lighter than the rest of the fields and the phase transition proceeds slowly leading to  strong signals in the stochastic gravitational wave spectrum.

\end{quote}

\vfill
\noindent\line(1,0){188}
{\scriptsize{ \\ E-mail:
\texttt{$^1$\href{mailto:aleksandr.azatov@NOSPAMsissa.it}{aleksandr.azatov@sissa.it}},
\texttt{$^2$\href{miguel.vanvlasselaer@NOSPAMsissa.it}{miguel.vanvlasselaer@sissa.it}}
}}

\newpage

\newpage
\section{Introduction}

	The nature of the Higgs boson remains one of the main unresolved puzzles of the modern particle physics. The discovery of the Higgs boson at LHC together with the null results in new physics searches provide no answers to this question. In particular it is well-known that the Higgs mass is quadratically sensitive to the new physics corrections which should naturally lift the Higgs boson mass to the scale of the cut-off, where UV completion is needed. This quadratic sensitivity of the mass of the Higgs boson, commonly dubbed as the \emph{hierarchy problem}, can be addressed in  models where the Higgs boson is a bound state of some new strong dynamics  (for reviews on the subject, see \cite{Hill:2002ap,Contino:2010rs,Bellazzini:2014yua}).  However it was soon realized that successful generation of the top quark mass together with absence of  flavour violating effects in the light quark sector require the strongly interacting system to be in the nearly conformal regime  \cite{Holdom:1981rm,Yamawaki:1985zg,Appelquist:1986an} (for a recent discussion of the problem, see \cite{Luty:2004ye,Rattazzi:2008pe}) for a significant  range of scales before it confines near the electroweak scale. This nearly conformal regime of very slow coupling evolution is often dubbed ``walking" in the technicolor literature.
 
It has been conjectured that the ``walking" regime appears when the $\beta$ functions have two complex poles (\cite{Kaplan:2009kr,Gorbenko:2018ncu}), with
the 
 imaginary part much smaller than the real one. 
The subject of this manuscript is a study of the phase transition (PT) from the deconfined to 
confined phase in the models with such ``walking" behavior.
Ideally we are interested in the theory with only fermions and gauge fields, where the
walking as well as the phase transition occur at strong couplings. Obviously in this case one
has to rely on nonpertubative techniques, e.g.,  the lattice simulations. 
The specific case of QCD with eight flavors,
$ N_f = 8, N_c = 3$, which is believed to be close to the exit of the conformal window
\cite{Appelquist:1996dq} has been studied on lattice \cite{Aoki:2013xza, Appelquist:2014zsa,Hasenfratz:2014rna,Appelquist:2016viq,Appelquist:2018yqe}  finding the remnants of the walking behavior, however the analysis of the phase transitions are still inconclusive. 
AdS/CFT duality  presents another avenue to address the problem \cite{Creminelli:2001th} (for
the recent studies, see \cite{Bruggisser:2018mus,Bruggisser:2018mrt,Baratella:2018pxi, Agashe:2019lhy,DelleRose:2019pgi}), however 
the analysis of the PT  was done only for the context of light dilaton .
 In this study, we instead analyze the phase transition in the
toy model proposed in \cite{Benini:2019dfy} where the walking and complex CFT occur at weak coupling. The
price for this perturbative regime is the presence of  scalar fields, which obviously do not
allow a solution for the hierarchy problem. However we believe that some qualitative features of
the phase transition that we find in this toy model will remain valid also in the realistic models
with no scalar fields.

If such PT  had occurred in the early history of the universe while it was  cooling down it might lead to very interesting phenomena. 
Particularly the holographic analysis of  models with near conformal dynamics  has shown that the transition is of the first order and generically  leads to the strong signal in stochastic gravitational wave  spectrum \cite{Randall:2006py,
Bunk:2017fic, vonHarling:2017yew, Baratella:2018pxi, Agashe:2019lhy,Konstandin:2011dr}. However the holographic results are valid only if the dilaton field is much lighter than the rest of the composite resonances. Our analysis on the other hand  can partially (we will show in the section \ref{sec:thermal} that we need as well a light dilaton in order to maintain the perturbativity of the model) relax this assumption and thus provide a very important complementary information.

The paper is organised as follows: In the section 2, we review the toy model of 
\cite{Benini:2019dfy} and discuss  the 
potential at tree  and one loop-level. In the section  3 we discuss the thermal corrections to the 
effective potential.  In the section 4, we discuss the phase transition and GW production and then we conclude.

\section{The perturbative walking model}
\label{sec:pwm}
We will consider the model proposed recently in \cite{Benini:2019dfy} as a toy model with \emph{Perturbative Walking Dynamics} (PWD).
The model is based on a  $SU(N_c)$-gauge theory with $N_c$ colors, $N_s$ complex scalar fields and $N_f$ fermions (Dirac) fields, both transforming in the fundamental representation of the gauge group. The model is governed by the Lagrangian\footnote{For perturbative analysis of this class of  models see also \cite{Antipin:2012kc,Hansen:2017pwe}.}
 \begin{equation}
 \mathcal{L} = -\frac{1}{4}F^{A}_{\mu\nu}F_{A}^{\mu\nu} + i\text{Tr}\bar{\psi} \slashed{D}\psi + \text{Tr} D_\mu \phi^{\dagger}D^{\mu}\phi - \tilde{h} \text{Tr}\phi^{\dagger}\phi\phi^{\dagger}\phi - \tilde{f} (\text{Tr}\phi^{\dagger}\phi)^2,
 \label{Lag}
 \end{equation}
  where the trace is  taken in the color-flavor space. As we anticipated in the introduction, our model contains scalar fields and thus it can not be considered as a realistic candidate to solve the Higgs hierarchy problem. However, its aim is to offer a perturbative, and consequently fully controllable, realization of the walking dynamics that allows for a quantitative study of the confinement-deconfinement phase transition. Let us do a quick summary of the ingredients at play: $2N_cN_s$ real scalar degrees of freedom (d.o.f.) (or $N_cN_s$ complex scalar d.o.f.), $N_c^2-1$ gauge bosons d.o.f. and $4N_cN_f$ fermionic d.o.f. (or $N_cN_f$ Dirac fields). The complete symmetry group is $SU(N_c)\times U(N_s)\times SU(N_f)^2\times U(1)$, with three couplings: the gauge coupling $g$ and  the self-couplings of the scalars for the double and the single trace , respectively $\tilde{f}$ and $\tilde{h}$. The renormalization group (RG) evolution of the system is thus
governed by three $\beta_i, i = {\lambda, \tilde{h}, \tilde{f}}$ functions which, in terms of the 't Hoof couplings
\bea
\lambda \equiv \frac{N_c g^2}{16\pi^2}, \qquad h \equiv \frac{N_c\tilde{h}}{16\pi^2}, \qquad f \equiv \frac{N_cN_s\tilde{f}}{16\pi^2}
\eea
 can be written as
 \bea 
 &&\beta_{\lambda} = -\frac{22-4x_s- x_f}{3}\lambda^2 + \lambda^3\Big(\frac{2}{3}(4x_s + 13x_f -34) -2\frac{x_s+x_f}{N_c^2}\Big),\nn
  &&\beta_{h} = 4(1+x_s)h^2+\frac{24}{N_cN_s}fh -  \Big(6-\frac{6}{N_c^2}\Big)\lambda h + \Big(\frac{3}{4}-\frac{3}{N_c^2}\Big)\lambda^2,\nn
 && \beta_f = 4\Big (1+\frac{4}{N_sN_c}\Big)f^2 + 8(1+x_s)fh + 12x_s h^2-\Big(6-\frac{6}{N_c^2}\Big)\lambda f + \frac{3x_s}{4}\Big(1+\frac{2}{N_c^2}\Big)\lambda^2.\nonumber\\
   \label{run3}
   \eea
   Analysing these equations becomes particularly simple in the Veneziano limit (that is to say, the limits $N_c, N_f, N_s \to \infty $ with $ x_s = \frac{N_s}{N_c}, x_f = \frac{N_f}{N_c}$ kept fixed). In this case, we can see that the equations for the 
    $\beta_\lambda=\beta_h=0$, become independent of $f$ and can be solved analytically leading to the two   Banks-Zaks\cite{Banks:1981nn,Caswell:1974gg}   perturbative  fixed points $(\lambda^\star, h^\star_+), (\lambda^\star, h^\star_-) $, characterized by the parameters $x_f, x_s$. The equation for the fixed point of the $f$ coupling, $\beta_f=0$, becomes
   \bea
\label{eq:poli}
    4 f^2 + \l[8(1+x_s)h^{\star} -6\lambda^{\star}\r]f + 12x_s h^{\star 2}+ \frac{3x_s}{4}\lambda^{\star 2}=0,
   \eea
where $\lambda^{\star}$ and $h^{\star}$ are the solutions of $\beta_\lambda=\beta_h=0$. 
Varying the parameters 
$x_s$ and $x_f$ we can make the solutions of \eqref{eq:poli}  complex or real.
In this simple setting, the walking behaviour occurs when the two couplings $h,\lambda$ satisfy the \emph{real} fixed-point condition and the roots of $f$-fixed point equation $\beta_f=0$ have a very small imaginary part. 
From \eqref{eq:poli}, we can see that the smallness of the imaginary part is controlled by $x_s$,  on the other hand $x_f$ enters only to set the order of magnitude of the couplings $\lambda \sim f \sim h$ via the Banks-Zaks condition (we develop on this point in the appendix \ref{rev-BIS}). 

	In particular, in this model the transition from the real to the complex fixed-points happens once $x_s$ crosses the critical value $\bar x_s=0.07039$ (for 
	details and generalisation to exact equations, see the appendix \ref{rev-BIS}). 
	Schematically, all along the walking regime, the RG equation for the coupling $f$ takes  the form\footnote{This type of scale separation is often dubbed Miransky's scaling \cite{Miransky:1984ef}.}
\bea
\frac{d f}{ d\ln \mu}=-A^2 -f^2
\label{beta:wal}\nonumber\\
\Rightarrow \log\l[\frac{\Lambda_{UV}}{\Lambda_{IR}}\r]\simeq \frac{\pi}{A}\sim \frac{1}{\sqrt{\beta_{walking}}}
\eea
As advocated in \cite{Kaplan:2009kr}, \eqref{beta:wal} is the typical form of $\beta$ function inducing walking behaviour. Indeed in this case the coupling $f$ 
remains approximately  constant (that is to say, the theory is almost conformal) 
for the range of scales between $[\Lambda_{IR},\Lambda_{UV}]$.
Solving the running equations numerically, we see that, once the system exits the 
walking regime, the coupling combination $f+h$ becomes negative, making the 
whole theory unstable. This triggers the development of the global and gauge 
symmetry breaking, which is an analogue of the confinement process in our toy 
model. In particular, once the $f+h$ becomes negative the vacuum develops a ``vacuum expectation value" (VEV) along the color-flavour-locking pattern direction
  \begin{equation}
  \label{eq:vev}
  \langle \phi^{a}_b\rangle = v \delta^{a}_{b},
  \end{equation} 
inducing the breaking of the gauge symmetry in the form $SU(N_c) \to SU(N_c-N_s)$. As a result, $N_s(2N_c-N_s)$ gauge fields and $N_s^2$ scalar d.o.f. obtain tree-level mass. The details are presented in the Table \ref{tab:table1}. 
\begin{table}[h!]
  \begin{center}
    \begin{tabular}{c|c|c|c}
      \textbf{Field} & \textbf{$\#$ of d.o.f} & {\bf mass}& {\bf mass at SB}\\
      \hline
      \multirow{3}{*}{Scalar} & $N_s^2-1$&  $2(3 \tilde{h}+N_s\tilde{f}) v^2$ &  $4 \tilde{h} v^2$\\ 
      & $2 N_c N_s-N_s^2$ & $2( \tilde{h}+N_s\tilde{f}) v^2$ &  0 \\ 
      &1& $6(\tilde h +N_s \tilde f) v^2$ &$\beta_{\tilde f+\tilde h/N_s} v^2$\\ 
      
      \hline
      \multirow{4}{*}{Vector} & $3(N_s^2-1)$&  $ g^2v^2 $& $ g^2v^2 $\\ 
      & $6 N_s( N_c-N_s)$ &  $\frac{1}{2} g^2v^2 $ &  $\frac{1}{2} g^2v^2 $\\ 
      &3& $(1-x_s) g^2 v^2$& $(1-x_s) g^2 v^2$\\ 
      &$3[(N_c-N_s)^2-1]$& $0$& $0$\\ %
      \hline
      \multirow{1}{*}{Fermion} & $4N_fN_c$&  $0$ &0  
    \end{tabular}
  \end{center}
    \caption{Field content and mass spectrum  of the model. In the second column, we indicate the number of degrees of freedom of each species, in the third we give the general form of the mass after symmetry breaking and in the fourth we give the mass expected at the  symmetry breaking scale. }
    \label{tab:table1}
\end{table}

At the instant of the symmetry breaking, the tree-level potential for the ``scalon'' field (that we can understand as the fluctuations along the VEV direction) vanishes, or equivalently phrased, becomes a flat direction. Loop-level corrections, taking the form of the well-known ``Coleman-Weinberg" (CW) potential\cite{Weinberg:1973am} (in $\overline{\rm MS}$ scheme), lift this flat direction:
 	\begin{equation}
\label{eq:cw}
 	V_{CW}(v) = g_i\frac{m_i^4(v)}{64\pi^2}\bigg[\log \Big(\frac{m_i^2(v)}{\mu_R^2}\Big)-c_i\bigg],
 	\end{equation}
where $g_i$ is the number of degrees of freedom of the species considered and $c_i = \frac{3}{2}\big(\frac{5}{2}\big)$ for bosons and fermions (vectors).
Note that during the phase transition process we will be exploring the potential in the regions far away from the renormalization point $\mu_R$.  To take the effects of the running of the coupling into account we will be using the RGE improved  CW effective potential. In the instance of  symmetry breaking (SB)
we can write the potential in the following compact way
\bea
&&f=-h \hbox{~~~at  SB}\nonumber\\
&&\beta_{f+h}^{SB}=8 x_s h^2 +\frac{3}{4}(x_s+1)\lambda^2\nonumber\\
&&V_{CW}\Big|_{\tilde{f}= -\frac{\tilde{h}}{N_s}}  =2 \pi^2 v^4 x_s\l[ 32 h^2 x_s +3 (x_s+1)\lambda^2\r]\log[v^2/\Lambda^2_{IR}],
\eea
 where we neglected sub-leading terms in $1/N_c$ expansion.  This potential is proportional to the $\beta_{f+h}$ function of the combined coupling $f+h$, which controls the spontaneous symmetry breaking. The loop-level lifting of the flat direction of the potential induces a non-vanishing mass for the \emph{scalon}, which we compute to be
\bea
m^2_{\text{scalon}}=\frac{16 \pi^2}{N_c N_s} \beta^{SB}_{f+h} v^2=\beta^{SB}_{\tilde f+\tilde h/N_s} v^2.
\eea
Let us note that, since the mass is controlled by the $\beta$ function during the symmetry breaking and not during the walking, it has just one-loop suppression compared to the masses of the other resonances\footnote{Remember that there is additional normalization factor $\frac{1}{\sqrt{2 N_s}}$ between the VEV $v$ and the scalon field in  order to have canonically normalized kinetic term.}. 
One can see that our construction  looks very similar to the usual Coleman-Weinberg scenario \cite{Weinberg:1973am} , where in the weak coupling case 
the large scale separation can also be generated, however  walking can strongly enhance this scale separation due to the factor $1/\sqrt{\beta_{walking}}$ as shown in   Eq.\ref{beta:wal} .

	Now, we would like to compare our results for the \emph{scalon} potential with the results obtained for the \emph{dilaton} in the models with spontaneous confinement transitions \cite{Baratella:2018pxi, Creminelli:2001th,Agashe:2019lhy}. This class of scenarios  are usually considered as partial  UV completions of the composite Higgs models inspired by the extra-dimensional Randall-Sundrum \cite{Randall:1999ee} (RS) models with Goldberger-Wise\cite{Goldberger:1999uk} (GW) radius stabilization. AdS/CFT duality relates them to strongly coupled, large N, approximate CFT models.
If the dilaton is the lightest degree of freedom, it will dominate the low energy potential,  which after integrating out the heavy species will  become (we use the notations of \cite{Agashe:2019lhy}):
\bea
V^{GW}_{\text{dilaton}}=\frac{N^2}{16\pi^2}\phi^4 \l[ \lambda_0+\lambda_0' g_{UV}\l(\frac{\phi}{\Lambda_{UV}}\r)^{\epsilon} \r],
\eea
 where the $\epsilon$ is the, very small, anomalous dimension of a almost marginal operator breaking the CFT. As a result, the scale of the spontaneous confinement is given by
\bea
\l<\phi\r>=\Lambda_{UV}\l(-\frac{1}{1+\epsilon/4}\frac{\lambda_0}{\lambda_0' g_{UV}}\r)^{1/\epsilon}. 
\eea
Again, the $UV/IR$ scale separation becomes
\bea
\Lambda_{IR}\sim  \Lambda_{UV}\l[O(1)\r]^{1/\epsilon}=
\Lambda_{UV}\l[O(1)\r]^{1/\beta_{walking}},
\eea
where, we see that the anomalous dimension $\epsilon$ controls the length of the walking. 

Thus the main differences compared to our perturbative model come from different $UV/IR$-scale separation as a function of the anomalous dimension of the operator breaking the CFT and, most importantly, the fact that, in the perturbative model, the $\beta$ function at the scale of symmetry breaking is much larger than its analogue during walking, $\beta_{SB}\gg \beta_{walking}$. Those differences are summarized in the table \ref{tab:wal}. 
\begin{table}[h!]
  \begin{center}
    \begin{tabular}{c|c|c}
    & PWD& RS with light dilaton\\
\hline
&&
\\
     scale separation $\Lambda_{UV}/\Lambda_{IR}$&$\sim O(1)^{\frac{1}{\sqrt{\beta_{walking}}}} $& $\sim O(1)^{\frac{1}{{\beta_{walking}}}} $\\
\hline
$\beta$ function at confinement &$\gg\beta_{walking}$&  $\sim \beta_{walking}$ 

    \end{tabular}
  \caption{Comparison between the perturbative walking dynamics (PWD) model with RS-like models of spontaneous confinement.}
    \label{tab:wal} 
\end{center}
\end{table}

At last, we would like to note that in our perturbative model  we can tune $\beta_{SB }$ to be small only at the price of making all the couplings very small, so that the loop suppression for the scalon mass becomes trivially important. We will see that this induces relatively long supercooling. 

\bfc
\includegraphics[scale=0.62]{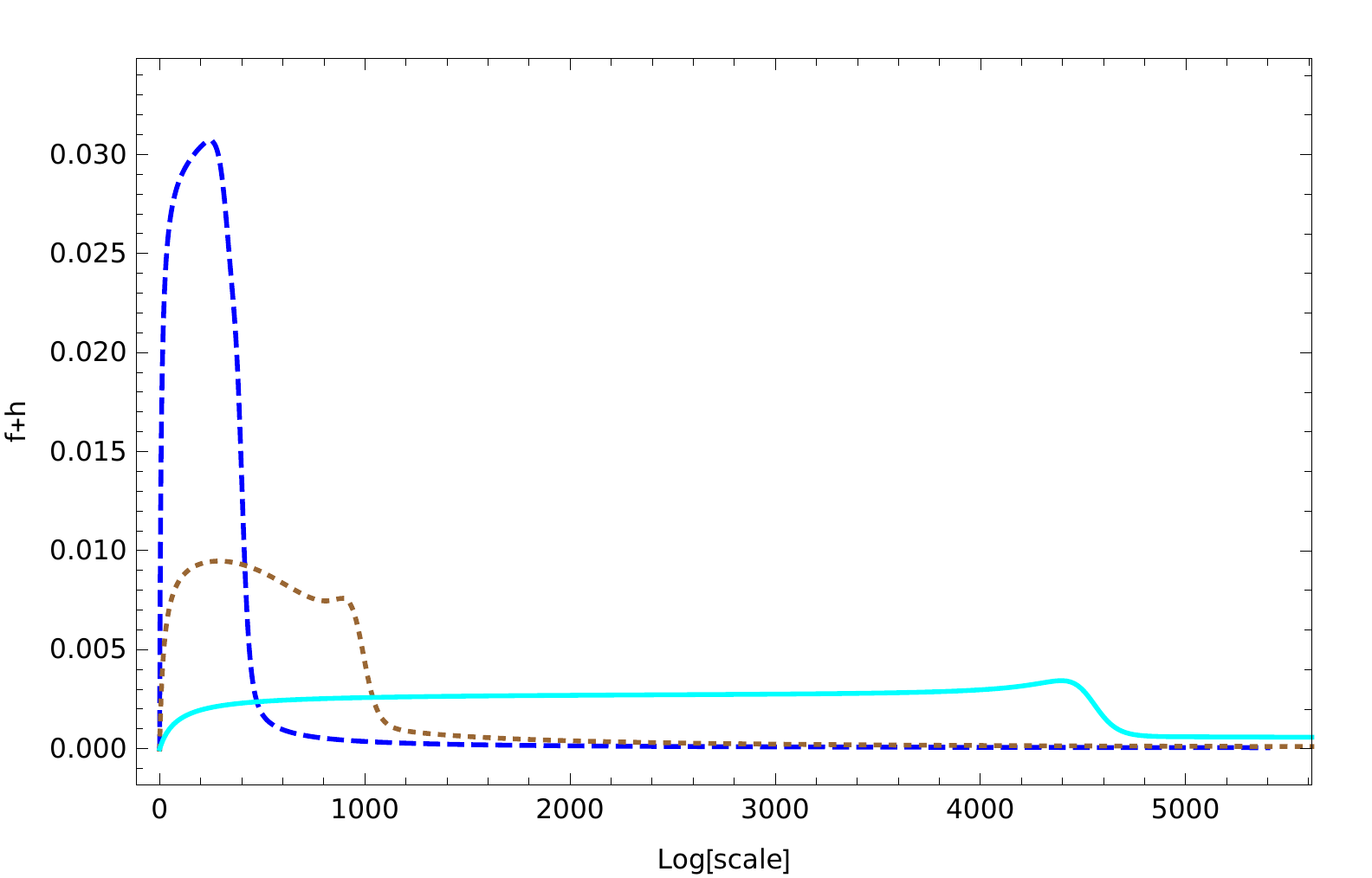}
\caption{\label{fig:walking} 
The dependence of the coupling $f+h$ on the scale for $N_s = 2, N_c = 25$. We have chosen the scale of symetry breaking to be equal to 1. The various curves from top to bottom correspond to $N_f=120,130,135$. We can clearly see that the length of the "walking plateau " increases once we go to the smaller values of the couplings.}
\efc

Now we can proceed to the numerical results for the couplings values and the parameters of the potential. In our case, the model develops a complex fixed-point only for the $N_s \geq 2$ (for $N_s=1$, the couplings $f, h$  become equal) and this imposes us to consider $N_c\sim \frac{N_s}{\bar x_s}\sim 25$ and, indeed, already for $N_c=25$, the walking behaviour starts to appear. The results for the $(h+f)$-coupling running are reported in the Figure~\ref{fig:walking}. To test the different regimes of the model, we will consider five sample points, keeping the same number of scalars and colors, but varying the number of fermions. All of those reference points (see table \ref{ref-values-run2} for the definitions) are required to lead to the UV free theory and to present a walking regime. 
The points are chosen in order to have various values of the $\beta_{SB}$ leading to phenomenologically very different  phase transitions once the temperature effects are taken into account. Note that the ``walking range" for all of those points is well beyond  what is needed for the Planck/weak hierarchy to be connected .

   \section{Thermal corrections to the potential}
   \label{sec:thermal} 	
 	To study the phase transition, we compute the potential at finite temperature  at one-loop order. As already mentioned above, to capture correctly the behaviour at the symmetry breaking scale, we set the tree-level potential to zero. It is well known that to account for the thermal excitations due to the temperature and the density in the early universe plasma, we have to add the thermal potential on the top of the zero temperature potential (see for example \cite{Curtin:2016urg}); 
 	\begin{equation}
 	V(T,m_i) = V_{CW}(m_i)+ V_T(m_i)
 	\end{equation}
 	where the Coleman-Weinberg   potential was defined in Eq.\ref{eq:cw} with the renormalization scale fixed to be 
 	\bea
 	\mu_R=w g,
 	\eea
 	and the thermal potential for bosonic fields part is given by:
 	\begin{equation}
 	V_T(m_i(v)) = \frac{g_i}{2\pi^2}T^4 J\Big(\frac{m_i^2(v)}{T^2}\Big), \qquad J(y^2) = \int \limits_0^{\infty} dx x^2\log \Big[1-\exp{(-\sqrt{x^2+y^2})}\Big]. 
 	\end{equation}
 	This function can be expanded in the limit of small and large argument using the following expansions \cite{Curtin:2016urg} (to save computation time, those are the mathematical expressions we use numerically)
 	\begin{align}
 	&J(y^2 \ll 1) = -\frac{\pi^4}{45}+\frac{\pi^2}{12}y^2-\frac{\pi}{6}y^3-\frac{y^4}{32}\log\bigg[\frac{y^2}{16\pi^2\exp[3/2-2\gamma]}\bigg],\nonumber \\
 	 & J(y^2\gg 1) = - \sum_{n = 1}^{m >3}\frac{1}{n^2}y^2K_2(y\cdot n).
 	\label{expansion_thermal}
 	\end{align}
 	where $\gamma \approx  0.5772156649$ is the \emph{Euler constant}, $m_i(v)$ is the mass of the particle i at the value $v$ of the VEV, $g_i$ the number of degrees of freedom of the considered fields and $K_2(z)$ are the second-kind Bessel function\footnote{{In our numerical calculations we have summed the Bessel functions up to $m=15$ and the matching between the low and high energy formulas was done for  $y^2=0.05$, in this way the differences with exact expressions were less than $\sim 0.01\%$. }}. To account for higher loops due to the Daisy diagrams at finite temperature, we can follow the so-called ``Truncated-Full-Dressing" procedure \cite{Curtin:2016urg}. Doing so, the full one-loop potential becomes 
 	\begin{equation}
 	\label{eq:potential}
 	V(v, T) = \sum_i V_{CW}(m_i^2+\Pi_i) + V_T(m_i^2+\Pi_i)
 	\end{equation}
 	where $\Pi_i$ are the so-called ``thermal masses", dependent on the VEV $v$ and the temperature for each degree of freedom. In our model, the expressions of the thermal masses read (see for example \cite{Comelli:1996vm})
 \begin{align}
 \label{eq:thermalmasses}
 &\Pi_s(T,N_c,N_s) = \frac{N_c^2-1}{2N_c} \frac{g^2T^2}{4}  + 2\Big(N_s-\frac{1}{N_s}\Big)\frac{\tilde{h}T^2}{12},\nonumber\\
&\Pi_{A, \text{Long}}(T, N_c,N_s, N_f) = \frac{1}{6}g^2T^2[2N_c + N_s+N_f] 
\nonumber\\
  &\Pi_{A, \text{Trans}} = 0.
\end{align}

The thermal corrections to the potential at high temperature make the origin $\phi=0$ the true minimum of the system and restore the broken symmetry. Once we consider the lowering of the temperature, the true minimum becomes defined by Eq.\ref{eq:vev} where  
$SU(N_c)$ symmetry is broken. At the same  time the thermal corrections insure that the 
second derivative at the origin is positive, thus the potential will have the two minima  separated 
by the potential barrier.
The  critical temperature $T_{\rm cr}$ is reached when both minima have the same energy.  Below this temperature there will be the first order phase transition, which will proceed either by quantum tunnelling or by thermal fluctuations.
The calculation of the tunnelling rates can be done numerically and we will discuss it in the next section.  As the model under consideration is only a toy model of the walking dynamics, it does not seem necessary to perform a full scan of the theory space. Instead we consider five reference points considering the minimalistic scenario with $N_c=25,N_s=2$. Note that the choice of the number fermions and scalars together with the requirement of the asymptotic freedom fix the values of the coupling at the scale of symmetry breaking. We report the values of the couplings at the exit point in the Table \ref{ref-values-run2}.

So far we have  assumed that  perturbative expansion of our theory is under control  once the couplings $\lambda, h,f \ll1$. However it is well-known that in thermal perturbation theory due to the IR effects the loop expansion \cite{Weinberg:1974hy,Arnold:1992rz}  becomes controlled 
by the $\sqrt{\hbox{coupling}}$. This  leads to additional constraints on the theory space where the perturbation theory is under control. We can estimate the loop expansion parameter by comparing the two and three loop corrections to the scalar mass. In particular comparing the $O(g^2)$ and $O(g^3)$ terms in the  $V_T(\hbox{A,{\rm Long}})$ we can see that thermal corrections are perturbative  for
\bea
\label{eq:thermalpert}
&&\frac{g \sqrt{3 N_c+ N_f/2+N_s/2} (1+N_s(N_c-2N_s))}{\pi (2 N_c N_s+N_s^2-3)}\sim  2.2 g \ll 1~~\Rightarrow\lambda\ll0.03,
\eea
and similarly for  $f, h\ll 0.03$.
Comparing this condition with the reference points given in the  Table \ref{ref-values-run2}, we can see that {\bf  P5} is under perturbative control and {\bf P1} is not perturbative and the rest of the points require more detailed analysis. So the results reported for them should be taken with some care.  Note that perturbative control of the thermal corrections to the potential push us towards  small values of the coupling constants and to the light scalon scenarios see discussion in Sec \ref{sec:pwm}.

Another issue regarding the perturbative treatment of the model is related to the fact that the couplings $\lambda, f,h$ have a Landau pole in the deep IR. This 
becomes particularly important since in order to study the phase transition we 
need to know the potential in the false vacuum at the origin of the potential $\phi=0$. We can cope with this by noting that the actual scale will be 
\bea
\Lambda(v,T,N) \equiv  \sqrt{ g^2 v^2 + \Pi_s T^2} 
\eea
so that perturbativity constraint on the running couplings $|h|,|\lambda|,|f| < 0.03 $ translates into the bound on the minimal temperature $T_{\rm min~pert}$ below which our analysis becomes inconsistent.

Having specified the potential including the thermal corrections and its range of validity we can proceed to the next step of calculating the nucleation rate.

\begin{table}[]
\begin{center}
\begin{tabular}{llllll}
\hline
\multicolumn{1}{|l|}{Ref. point} & \multicolumn{1}{l|}{\bf P1} & \multicolumn{1}{l|}{\bf P2} & \multicolumn{1}{l|}{\bf  P3} & \multicolumn{1}{l|}{\bf P4}  & \multicolumn{1}{l|}{\bf P5}    \\ \hline
\multicolumn{1}{|l|}{$N_f$} & \multicolumn{1}{l|}{120} & \multicolumn{1}{l|}{130}& \multicolumn{1}{l|}{133}  & \multicolumn{1}{l|}{134} & \multicolumn{1}{l|}{136} \\
\hline
\multicolumn{1}{|l|}{$\epsilon$} & \multicolumn{1}{l|}{0.0362} & \multicolumn{1}{l|}{0.0149} & \multicolumn{1}{l|}{0.0064} & \multicolumn{1}{l|}{0.00426} & \multicolumn{1}{l|}{0.00213}  \\ \hline
\multicolumn{6}{|l|}{ ~~~~~~~~~~~~~~~~~~~~~~~~~~$N_c=25$,~~~~~~~~~~~~~~~~~$N_s=2$}                                                                                               \\
\hline
\multicolumn{1}{|l|}{$\lambda$ at SB} & \multicolumn{1}{l|}{0.0473} & \multicolumn{1}{l|}{0.0166} & \multicolumn{1}{l|}{0.009} & \multicolumn{1}{l|}{0.0067} & \multicolumn{1}{l|}{0.0021} \\ \hline
\multicolumn{1}{|l|}{naive loop expansion} & \multicolumn{1}{l|}{$\sim$1.2} & \multicolumn{1}{l|}{$\sim $0.75} & \multicolumn{1}{l|}{$\sim $0.5} & \multicolumn{1}{l|}{
$\sim 0.45 $} & \multicolumn{1}{l|}{$\sim$0.3} \\ \hline
\multicolumn{1}{|l|}{$h$ at SB} & \multicolumn{1}{l|}{0.066} & \multicolumn{1}{l|}{0.023} & \multicolumn{1}{l|}{0.0126}& \multicolumn{1}{l|}{0.0093} & \multicolumn{1}{l|}{0.003} \\ \hline
\multicolumn{1}{|l|}{$f$ at SB} & \multicolumn{1}{l|}{-0.066} & \multicolumn{1}{l|}{-0.023} & \multicolumn{1}{l|}{-0.0126}& \multicolumn{1}{l|}{-0.0093} & \multicolumn{1}{l|}{-0.003} \\ \hline
%
%
\multicolumn{1}{|l|}{$T_{\text{min pert}}/w$ } & \multicolumn{1}{l|}{$- $} & \multicolumn{1}{l|}{$2\times 10^{-4} $}& \multicolumn{1}{l|}{$<10^{-10}$} & \multicolumn{1}{l|}{$<10^{-10}$} & \multicolumn{1}{l|}{$<10^{-10}$}\\ \hline
\multicolumn{1}{|l|}{$T_{\text{cr}}/w$ } & \multicolumn{1}{l|}{0.3} & \multicolumn{1}{l|}{0.167}& \multicolumn{1}{l|}{0.116} & \multicolumn{1}{l|}{0.096} & \multicolumn{1}{l|}{0.052}  \\ \hline
\multicolumn{1}{|l|}{$\frac{m^2_{\text{scalon}}}{m^2_{\text{gauge}}}$ at SB} & \multicolumn{1}{l|}{0.049} & \multicolumn{1}{l|}{0.017}& \multicolumn{1}{l|}{0.009} & \multicolumn{1}{l|}{0.0069} & \multicolumn{1}{l|}{0.0022}  \\ \hline
\multicolumn{6}{|l|}{ } \\
\multicolumn{6}{|l|}{ ~~~~~~~~~~~~~~~~~Phase transition parameters for $w=10^5$ GeV}                                                                                               \\
\multicolumn{6}{|l|}{\hspace{6 cm}{\bf \Large $\Downarrow$}}                                                                                               \\
\hline
\multicolumn{1}{|l|}{$T^{\text{nuc}}$ } & \multicolumn{1}{l|}{0.145} & \multicolumn{1}{l|}{0.0069}& \multicolumn{1}{l|}{{2.2 $\times 10^{-4}$}}  & \multicolumn{1}{l|}{$1.05\times 10^{-5}$} & \multicolumn{1}{l|}{$-$} \\ \hline
\multicolumn{1}{|l|}{$T^{\text{per}}$ } & \multicolumn{1}{l|}{0.14} & \multicolumn{1}{l|}{0.0066}& \multicolumn{1}{l|}{ {2.1 $\times 10^{-4}$}}  & \multicolumn{1}{l|}{$1.04\times 10^{-5}$} & \multicolumn{1}{l|}{$-$} \\ \hline
\multicolumn{1}{|l|}{$\alpha$ } & \multicolumn{1}{l|}{0.042} & \multicolumn{1}{l|}{960}& \multicolumn{1}{l|}{$3\times 10^{8}$} & \multicolumn{1}{l|}{$3\times 10^{13}$} & \multicolumn{1}{l|}{$-$}  \\ \hline
\multicolumn{1}{|l|}{$\beta/H=T\frac{d}{dT}\l(\frac{S_3}{T}\r)$ } & \multicolumn{1}{l|}{485} & \multicolumn{1}{l|}{377} & \multicolumn{1}{l|}{{350}}& \multicolumn{1}{l|}{{340}} & \multicolumn{1}{l|}{$-$}\\ \hline
\multicolumn{1}{|l|}{$\alpha_\infty$ } & \multicolumn{1}{l|}{0.05} & \multicolumn{1}{l|}{9}& \multicolumn{1}{l|}{$5400$} & \multicolumn{1}{l|}{$1.4\times 10^{6}$} & \multicolumn{1}{l|}{$-$}  \\ \hline
\end{tabular}
\end{center}
\caption{\label{ref-values-run2}
The couplings and the phase transition parameters for the five reference points.
}
\end{table}

\section{Phase transition in PWD}
\bfc
\includegraphics[scale=0.6]{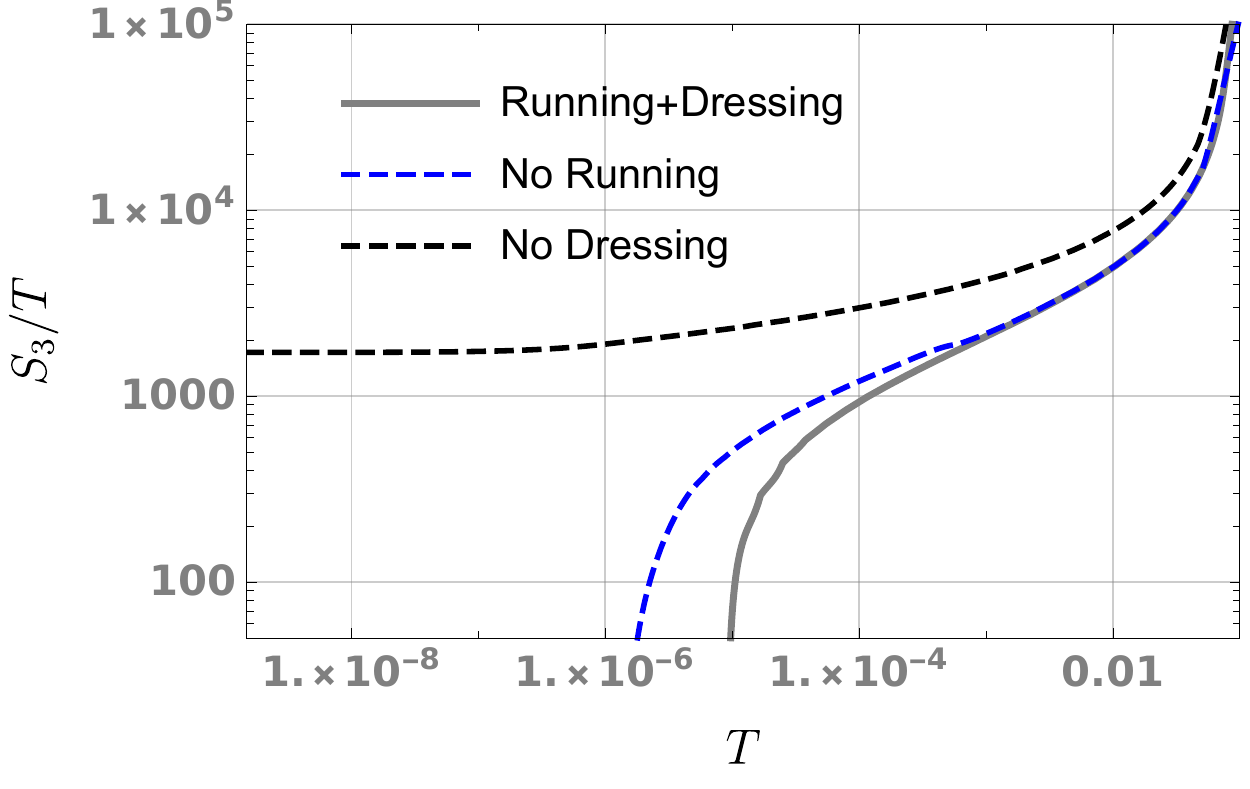}
\includegraphics[scale=0.6]{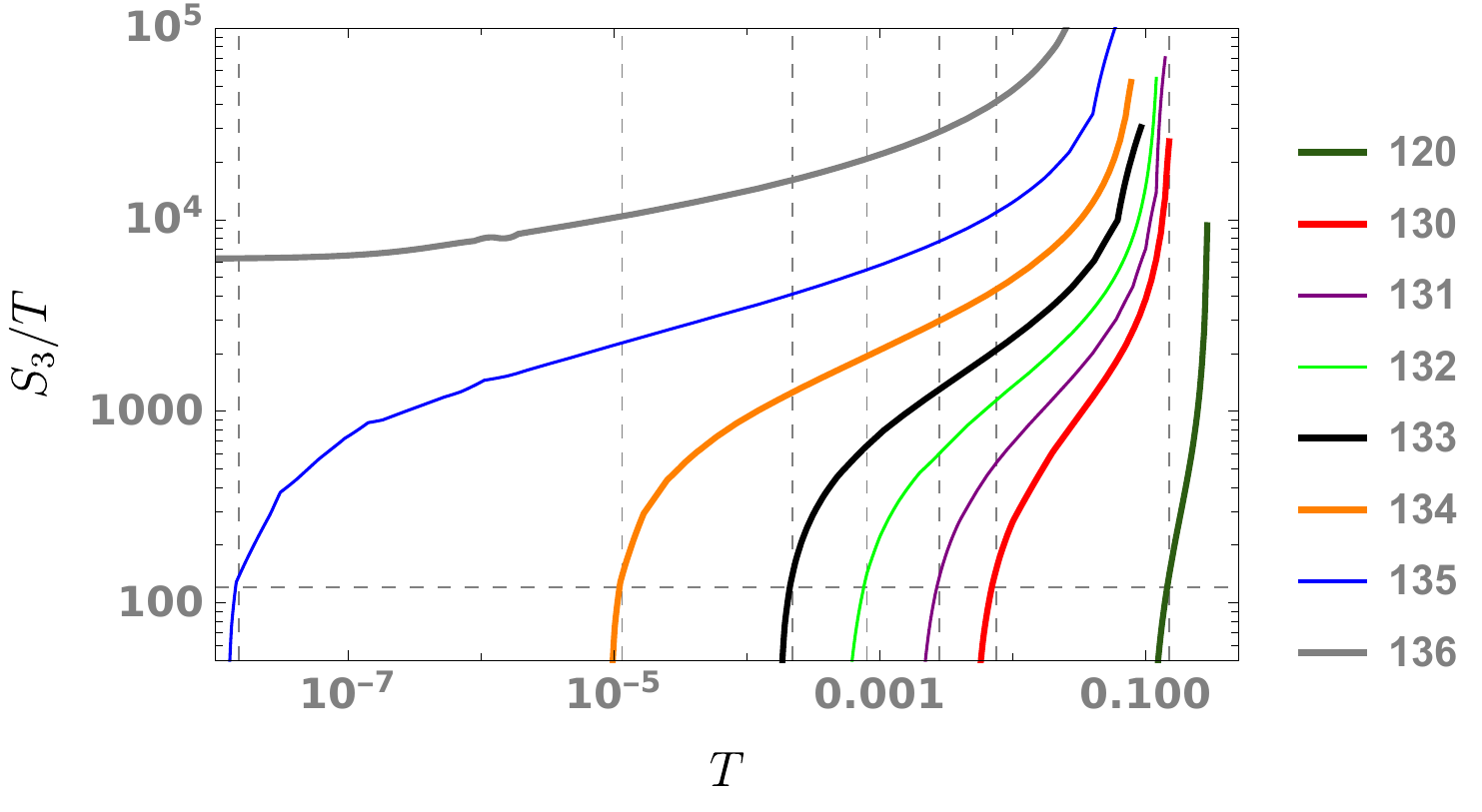}
\caption{On the left, illustration of the effects of the running and of the thermal masses for $N_f = 134$. We can see that the effect of the running, as it increases the couplings, is to shorten the supercooling, but it remains slight though. On the other hand, the effect of the thermal masses is quite dramatic and determines the moment of the transition. On the right, we represent the $\frac{S_3(T)}{T}$ as a function of the temperature for $N_f = 120, 130-136$.  The horizontal dotted line represent the nucleation condition  $\frac{S_3(T)}{T} \approx 120$. We can check in the table that this estimate is rather precise. }
\label{ThermalMasses}
\efc
\bfc
\includegraphics[scale=0.55]{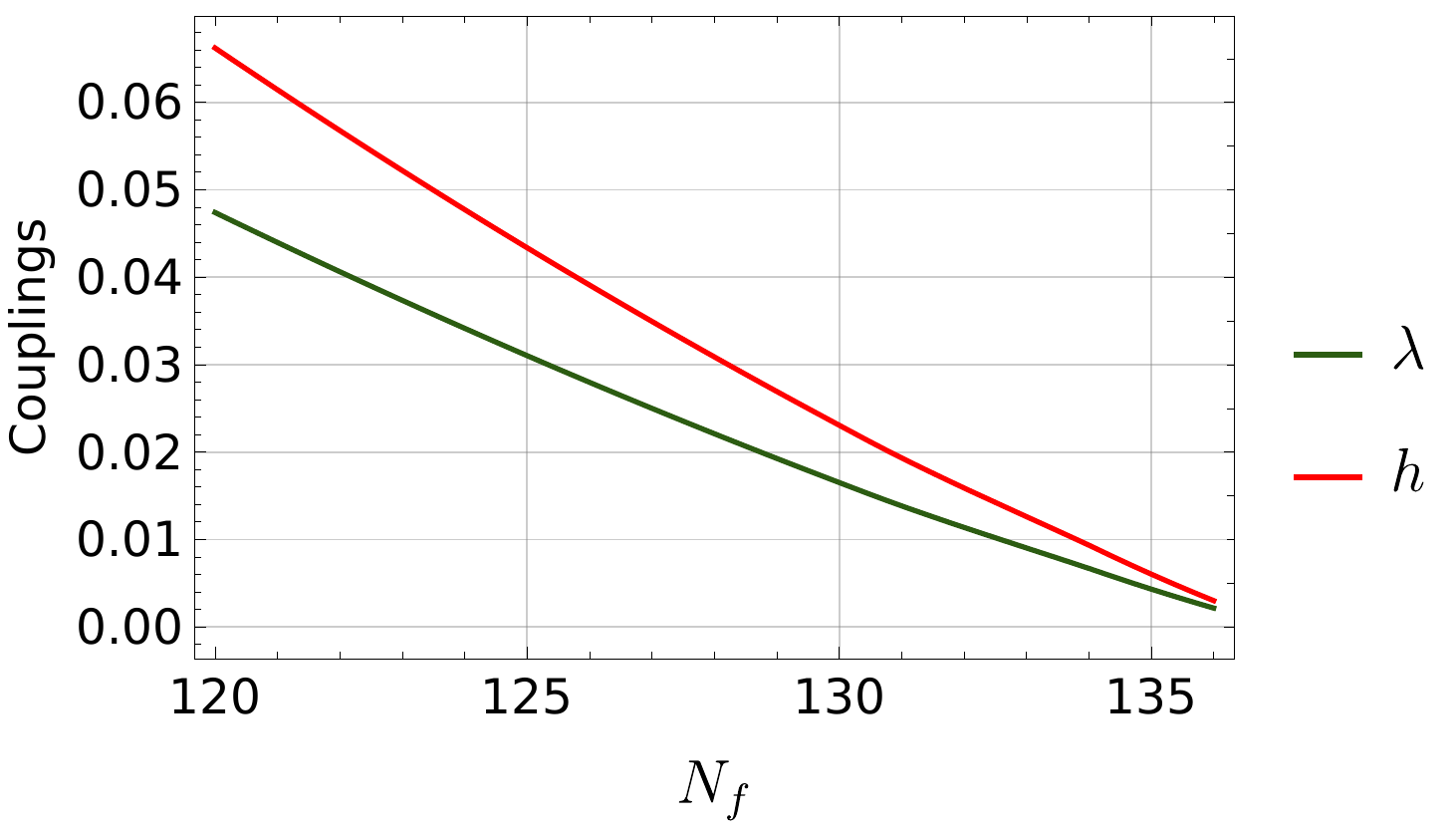}
\includegraphics[scale=0.58]{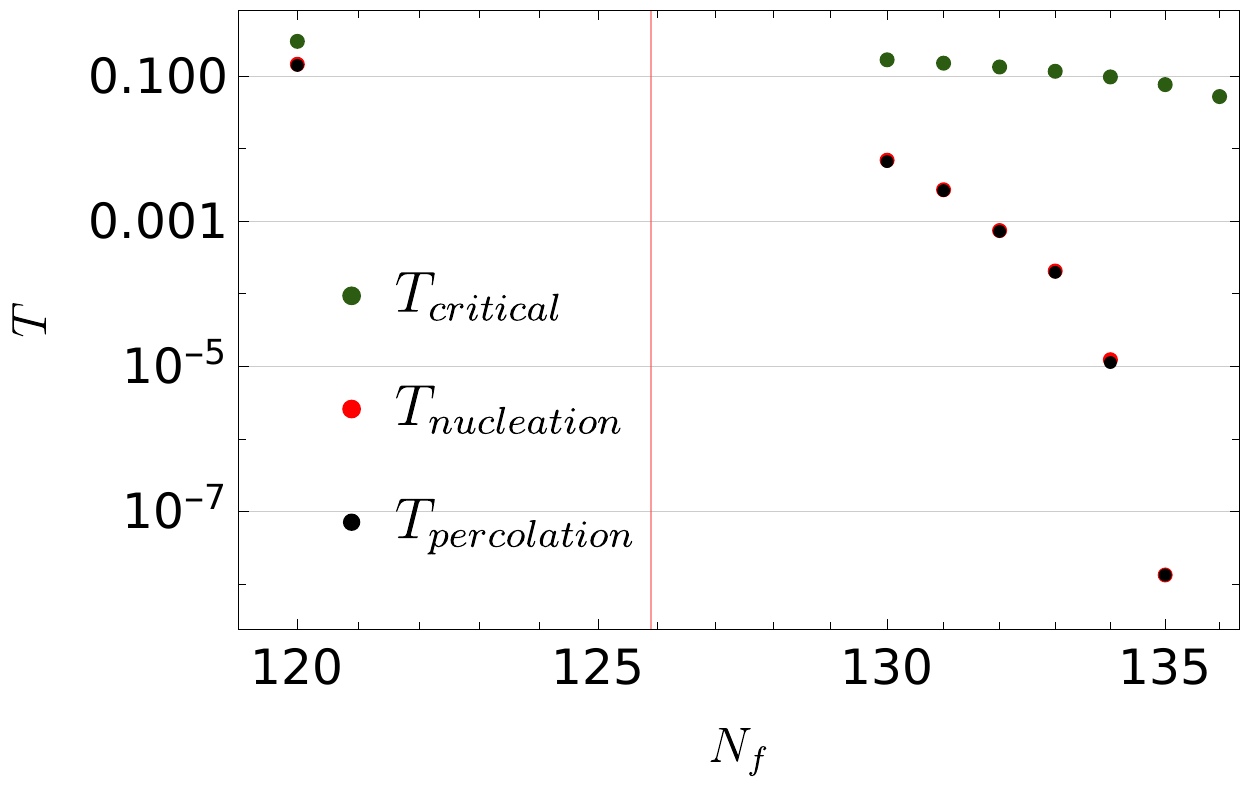}
\caption{We represent, on the left, the values of the coupling $\lambda$ and $h$ at the scale of symmetry breaking as a function of the number of fermions and, on the right, the values of the critical temperature, nucleation temperature and percolation temperature as a function of the number of fermions. The red line denotes the approximate number of fermions for which we expect the loop series expansion to break.
\label{fig:couplingtper}
}
\efc
\bfc
\includegraphics[scale=0.66]{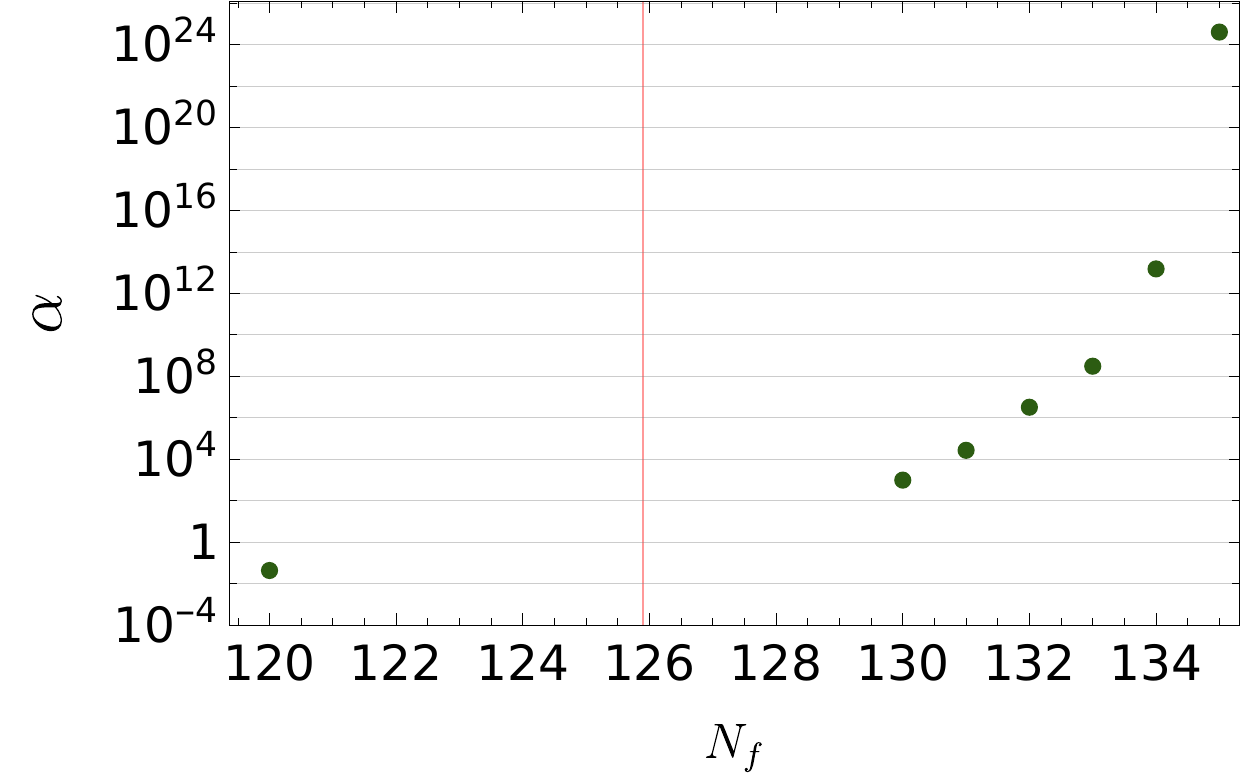}
\includegraphics[scale=0.66]{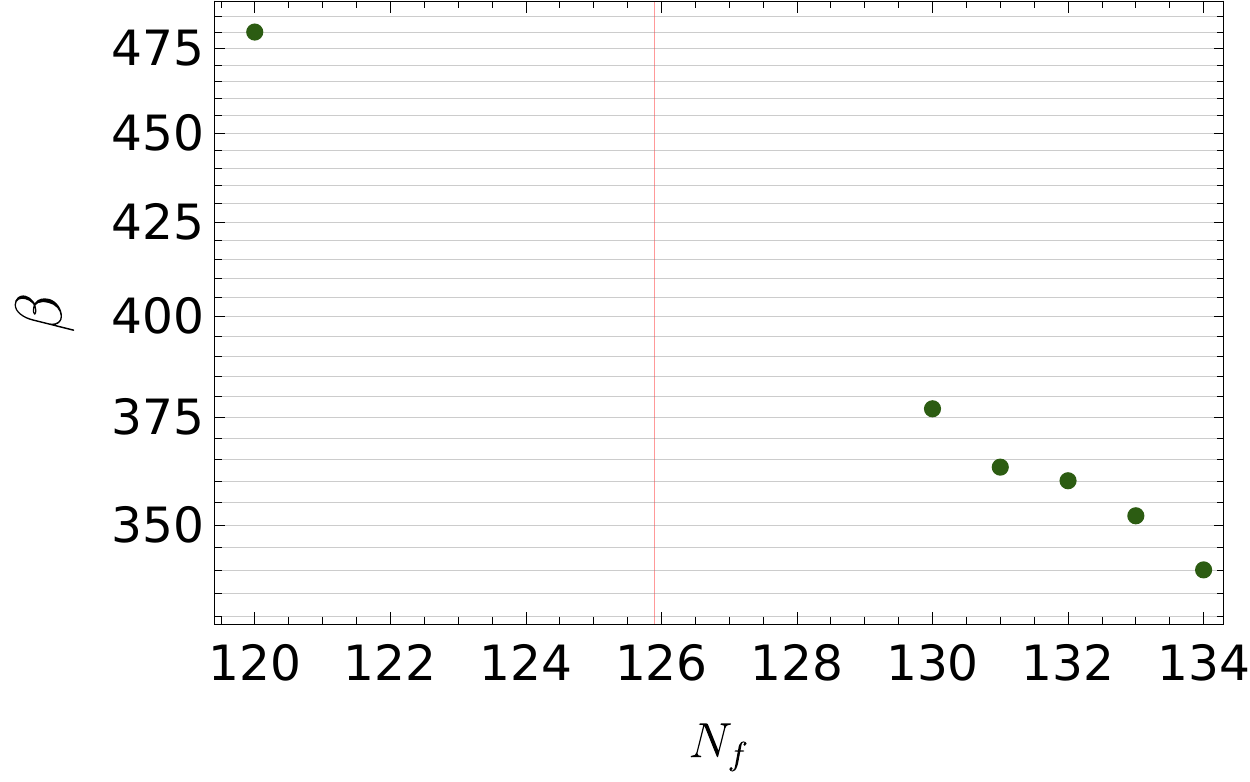}
\caption{We represent, on the left the values of the normalized energy budget $\alpha(T_p)$ and, on the right, the velocity of the transition $\beta$ at the time of percolation as a function of the number of fermions. 
\label{fig:alphabeta}}
\efc
In the previous section we have argued that at temperatures  below $T_{\text{cr}}$   the false and  true vacuum  are separated by a  potential barrier, so the phase transition will occur either by thermal fluctuation or by quantum tunnelling and will be first order. The probability of the transition (called "nucleation rate") can be easily calculated using the \emph{bounce action}:
\bea
\Gamma(T)\sim \text{max} \l[T^4 \l(\frac{S_3}{2\pi T}\r)^{3/2}\text{Exp}(-S_3/T), R_0^{-4} \l(\frac{S_4}{2\pi}\r)^2 \text{Exp}(-S_4)\r]
\eea
where $S_3,S_4$ are the usual action of $O(3),O(4)$ symmetric bounces (we find numerically that $S_3$ bounce is always dominating). The solution for the bounce can be found numerically using the shooting method and in spite of the 
multitude of the fields in our system the tunnelling will occur along the scalon 
direction, which simplifies drastically the calculation (see discussion in the 
appendix \ref{sec:1dbounce}). One subtlety is that nucleation should occur at the 
temperature above $T_{\rm min~ pert}$, since for  the temperatures below it the 
potential at the origin becomes non-perturbative. 
Due to the absence of UV masses and the fact that the Coleman-Weinberg 
potential is almost scale invariant, we expect the function $\frac{S_3}{T}$ to be 
nearly scale invariant as well up to the $\log T$ corrections. Resummation of the 
hard thermal loops (see Eq. \ref{eq:potential}) modifies this behavior  making the transition  faster.  One of the peculiarities of the PWD models is that we have a large number of fermion fields charged under the gauge group but not coupled to the scalar fields, as a result increasing the thermal masses of the gauge fields.
(see Eq.\ref{eq:thermalmasses}). Effect of these resummation is shown on the 
Fig.\ref{ThermalMasses}, where we have plotted the $O(3)$ bounce action for the potential with and without ``Truncated-Full-Dressing'' procedure.  We can see that thermal resummation accelerated the phase transition. On top of this we have indicated the effect of the running of the coupling constants which turns out to be subdominant (see for a recent discussion of the running effects \cite{DelleRose:2019pgi}).

The moment of transition can be estimated  by equating the  nucleation rate to the Hubble expansion at the moment
\bea
\label{eq:nucleation}
&&\Gamma(T_*)= H^4(T_*),\nonumber\\
&&H^2\equiv \frac{\rho_{\text{rad}}+\rho_{\text{vac}}}{3 M_{pl}^2}= \frac{1}{3 M_{\text{pl}}^2}\l(\frac{\pi^2 g_*}{30}T^4+\Delta V \r),
\eea
with $M_{\text{pl}} \equiv 2.435 \times 10^{18}$ the reduced Planck mass. 
This procedure defines the nucleation temperature denoted by $T^{\text{nuc}}$ in the table \ref{ref-values-run2} and Fig. \ref{fig:couplingtper}. When the energy budget of the universe is dominated by relativistic species energy, a simple estimate of the nucleation is given by the hierarchy created between the Planck scale and the scale of symmetry breaking
\begin{equation}
\frac{S_3(T^{\text{nuc}})}{T^{\text{nuc}}} \approx 4\log\bigg[\frac{T^{\text{nuc}}}{H(T^{\text{nuc}})}\bigg] \approx 4\log\bigg[\frac{M_{pl}}{T^{\text{nuc}}}\bigg] +... \sim 120.
\label{estimateTn}
\end{equation}
This very rough estimate provides the values which are close to the exact solutions
of Eq.~\ref{eq:nucleation} due to the  fast variation of the  quantity $\frac{S_3(T)}{T^{}}$ with the temperature, which controls the nucleation rate.
More precisely the temperature of the phase transition  can be found  by following the   procedure outlined in \cite{Ellis:2018mja}  
\bea
I(T_p)\gtrsim 0.34, ~I(T)=\frac{4 \pi}{3}\int_{T_p}^{T_{cr}}\frac{d T' \Gamma(T')}{T'^4 H(T')}\l[v_w\int_T^{T'} \frac{d   \tilde T}{H(\tilde T)}\r]^3
\eea
where the condition $I(T_p)\sim 0.34$ implies that the false 
vacuum occupies less than $\text{Exp}[-I(T_p)]\sim70\%$ of  the 
total space of the universe. This temperature is referred as the 
percolation temperature.  An accurate calculation of the integrals 
requires the knowledge of the bubble expansion velocity $v_w$ which is fixed by the equilibrium of the potential difference 
between the true and false vacuum and the pressure due to the friction force. The expression for the friction pressure are particularly simple in the relativistic wall case \cite{Dine:1992wr,Bodeker:2009qy,Bodeker:2017cim}
\bea
\Delta{\cal P}_{LO}\to\frac{T^2}{24} \sum  \Delta m^2,~~~	\Delta \mathcal{P}_{\text{NLO}} \sim T^3\gamma g^3 \frac{\sum \Delta m}{16\pi^2} ,
\eea
where LO,NLO stand for leading order and next to leading order effects and $\gamma$ is the Lorentz factor. For our reference points we find that only for {\bf P1 }the LO  pressure can balance the driving force due to the potential difference and even in this case both are very close to each 
other, so that we can assume $v_w\sim 1$.
\bea
\label{eq:vel}
\l.\baa{c}
\hbox{\bf P1} :~~~ \Delta V\sim \Delta{\cal P}_{LO}\\
\hbox{\bf P2-P4  } :~~~ \Delta V > \Delta{\cal P}_{LO}\\
\eaa\r \} \Rightarrow v_w\simeq 1
\eea
The other important parameters
characterising the phase transition are: the energy available to the transition $\alpha(T)$ and the speed of the transition $\tilde\beta(T)$. They are defined as follows:
\bea
\alpha=\frac{\Delta V-\frac{T}{4} \frac{\d \Delta V}{\d T}}{\rho_{\text{rad}}},~~~\tilde{\beta} \equiv \frac{\beta}{H}= -\frac{d(S_3/T)}{H dt} = T\frac{d}{d T}\l(\frac{S_3}{T}\r).
\eea
The $\alpha$ parameter, the latent heat of the transition normalized by the radiation energy of the universe, is related to the amount of supercooling in the following sense: as the temperature decreases below the critical temperature, the difference of depth between the true and the false vacua $\Delta V$ increases, while the relativistic energy gets redshifted by the expansion of the universe. The $\Delta V$ gives an order of magnitude estimate of the energy liberated by the nucleation and transferred to heating the plasma and accelerating the wall of the bubble\cite{Espinosa:2010hh}. Thus, a larger supercooling induces a larger energy budget parameter $\alpha$.
%
The normalized speed of nucleation $\tilde{\beta} \sim \frac{t_{\text{expansion}}}{t_{\text{transition}}}$, with $t_{\text{transition}}$ the typical time the transition takes to complete and $t_{\text{expansion}}$ the Hubble time, measures how fast a bubble nucleates with respect to the expansion of the universe, giving an estimated of the speed of completion of the transition.
The numerical values of all these parameters are reported in the table \ref{ref-values-run2}, where we have set the scale of the model to be $w=10^5$ GeV (the results for the other values of the scale are reported in the appendix \ref{sec:3vaiousw}).  

At last we would like to comment that  the reference     point {\bf P5}   will 
never  satisfy the nucleation condition (Fig.\ref{ThermalMasses}), and the system 
will remain trapped in the false vacuum. The situation is very similar to the 
holographic models with very light dilaton \cite{Creminelli:2001th}. One 
possibility, which was advocated \cite{Baratella:2018pxi,vonHarling:2017yew}, is 
that QCD confinement can trigger the phase transition in this case.  We will not 
analyze further this possibility in this paper. Note that this similarity is not 
coincidental and   comes from the fact that for {\bf P5} we have very small 
couplings, thus light dilaton in the spectrum, so that analysis of   
\cite{Baratella:2018pxi,vonHarling:2017yew}  are applicable to PWD as well.

\subsection{GW signal in the toy example} 

 In the previous sections, we have determined the typical range of parameters in which we expect the transition to be a FOPT. It is well known that, due to the out-of-equilibrium nature of the domain wall, a FOPT happening in the plasma of the early universe is expected to produce a stochastic gravitational wave signal. In this section, we review quickly the physics of the emission of gravitational waves emission during FOPT and present the typical spectrum predicted by the different points we singled out above. 
 
 	Three main contributions to the GW waves signal have been determined so far: the \emph{scalar field} contribution, originating from the collision of the bubbles, a \emph{sound waves} contribution coming from sound waves propagating into the plasma, and a \emph{turbulent} contribution due to turbulent motion. Following the recommendations of \cite{Caprini:2019egz}, we will ignore the turbulence contribution due to the large uncertainties and, in our computation of the spectrum, will focus only on the ``sound waves" and ``bubble collision" contributions. We already introduced the $\alpha$ parameter giving an estimate of the energy available to the transition and the $\beta$ parameter providing its velocity. Before to enter the physics of the different contributions, let us introduce two other important quantities entering into the computation of the GW signal emitted, the {reheating temperature} and the energy distribution  between  the motion of the wall and the excitation of the plasma. 
 	
	Immediately after the transition, we expect a  reheating to happen, bringing a correction to the Hubble constant. 	 As a consequence, we compute the $T_{\text{reh}}$, which is the temperature immediately after the phase transition completed, via the conservation of energy relation
\bea
{{(1-\Omega_{GW})(|\Delta V| + \rho_{\text{rad}}|_{T =  T_p}) = \rho_{\text{rad}}|_{T = T_{\text{reh}}}}}\nonumber
\\
\Rightarrow T_{\text{reh}}\approx (1+\alpha)^{1/4} T_p
\eea
where we neglected the energy going to the gravitational waves.
 Then, as the transition releases energy, we need to know which fraction of this energy goes into accelerating the wall and which fraction goes to the plasma kinetic energy, via the friction. From energy conservation consideration and to formalize this separation, we define  two parameters
 \begin{equation}
 \kappa_{\text{wall}}, \qquad \kappa_{\text{fluid}} = 1 - \kappa_{\text{wall}},
 \label{parameters}
 \end{equation}
$\kappa_{\text{wall}}$ is a measure of the ratio of energy going to the wall kinetic energy
\begin{equation}
\kappa_{\text{wall}} \equiv \frac{E_{\text{wall}}}{E_{\text{total}}}
\end{equation}

For the reference points  {\bf P1-P4} the wall always expands relativistically, however the $k_{\rm wall}$ becomes vanishingly small as soon as the terminal velocity is reached since the portion of the energy stored in the wall starts to decrease as inverse of the  bubble  radius. In order to understand whether the terminal velocity will be reached one can look at 
\bea
\Delta \mathcal{P}_{\text{NLO}}^{max} \sim T^3\gamma^{\rm collision} g^3 \frac{\sum \Delta m}{16\pi^2} \sim T^3 g^3 \frac{\sum \Delta m}{16\pi^2} \times \l(\frac{R^{\rm collision}}{R_c}\r),
\eea
where $R_c$ is the radius of the bubble at the instance of nucleation and can be estimated either  $R_c \sim \Big(\frac{3}{2\pi}\frac{S_3}{\Delta V}\Big)^{1/3}$  \cite{Ellis:2018mja}  or directly  numerically from  the profile of the  bounce solution. We find that our reference points fell into {three} categories
\begin{enumerate}
	\item {\bf P1-P2: relativistic with terminal velocity} 
	\newline 
In this case only the sound waves  are important and the energy will be distributed as follows:
\begin{equation}
 \kappa_{\text{wall}} =0, \qquad \kappa_{\text{fluid}} =1.
 \label{parameters2}
 \end{equation}

 \item 
 {\bf P3-P4: Runaway Regime} 
\newline
The release of energy is large enough to overcome all the source of friction and then the wall keeps accelerating until the collision. Mathematically, the condition writes
\bea 
 \Delta V>\big(\Delta \mathcal{P}_{\text{LO}}+\Delta \mathcal{P}^{max}_{\text{NLO}}\big) 
\eea 
In this case, the parameters introduced above become
	\begin{equation}
 \kappa_{\text{wall}} = 1 - \frac{\alpha_{\infty}}{\alpha}, \qquad \kappa_{\text{fluid}} = 1 - \kappa_{\text{wall}},~~~ \alpha_{\infty} = \frac{\mathcal{P}_{\text{LO}}}{\rho_{\text{radiation}}}. 
 \label{parameters3}
 \end{equation}
One can see that both sound waves as well as the bubble collisions are important for the generation of the gravitational waves.

\item {\bf P5:  } Trapped in the false vacuum, unless some other effect can trigger the PT.
\end{enumerate}

With all those quantities in hands, we can now go to the computation of GW spectra. 
\begin{itemize}
\item 
The first contribution is the so-called \emph{scalar field} contribution. During the phase transition, at the junction between the two phases, the VEV of the scalars involved in the transition smoothly interpolates between the two phases. The gradient in those background fields induces shear stresses. 
	The most recent numerical computation of the spectrum generated by this process can be approximated by  \cite{Cutting:2018tjt}
\begin{equation}
\frac{d\Omega_{\phi}h^2}{d\text{ln}(f)} = 4.7\times 10^{-8}\bigg(\frac{100}{g_\star}\bigg)^{1/3}(H_{\text{reh}}R_\star)^2\bigg(\frac{\kappa_{\text{wall}}\alpha }{1+\alpha}\bigg)^2S_{\text{wall}}(f, \tilde{f}_\phi)
\end{equation}
where $g_\star$ is the number of relativistic degrees of freedom, $\kappa_{\text{wall}}$ is the fraction of kinetic energy stored in the motion of the wall, $H_{\text{reh}}$ is the Hubble constant evaluated at the reheating temperature and $R_\star$ is the size of the bubble at the collision. The numerical fit to the  spectral function reads 
\begin{equation}
S_{\text{wall}}(f,\tilde{f}) = \frac{(a+b)^c \tilde{f}^bf^a}{\big(b\tilde{f}^{\frac{a+b}{c}}+a f^{\frac{a+b}{c}}\big)^c} \qquad a =3, \quad  b = 1.51, \quad c = 2.18,
\end{equation}
with peak frequency 
\begin{equation}
\tilde{f}_\phi = 16.5 \times 10^{-5}\bigg(\frac{T_{reh}}{100} \bigg)\bigg( \frac{g_\star}{100}\bigg)^{1/6}\bigg(\frac{3.2}{2\pi R_\star} \frac{1}{H_{reh}}\bigg) \text{ Hz}.
\end{equation}
and the typical bubble radius can be estimated to be
\bea
R_*=\frac{(8\pi)^{1/3}v}{\beta}. 
\eea

\item Another important mechanism of  gravitational wave production comes from the sound waves in the plasma. In this case the spectrum of the stochastic gravitational wave background can be estimated following the recent recommendations in \cite{Caprini:2019egz}
 	\begin{equation}
 	\frac{d\Omega_{gw,0}h^2}{d \text{ln}(f)} = 
 	\begin{cases}
    0.678 h^2 F_{gw,0}K^2(H_{\text{reh}} R_\star/c_s)\tilde{\Omega}_{gw,0}C(f/f_{p,0}),& \text{if } \frac{H_{\text{reh}}R_\star}{K^{1/2}}> 1\\
    0.678h^2 F_{gw,0}K^{3/2}(H_{\text{reh}} R_\star/c_s)^2\tilde{\Omega}_{gw,0}C(f/f_{p,0}),              & \text{if }   \frac{H_{\text{reh}}R_\star}{K^{1/2}}< 1
\end{cases}
 	\end{equation} 
and the two regimes in the equation above correspond to the time scale of the shock formation being larger or smaller than the corresponding 
 Hubble time, $\tau_{sh} >,< \frac{1}{H}$.  The sound wave production efficiency
 is given by \cite{Espinosa:2010hh}
 \bea
 K\approx \frac{3}{4}\frac{k_{sw}\alpha}{(1+\alpha)} ,~~~
    \kappa_{sw} = 
    \kappa_{\rm fluid}\times \frac{\alpha_{}}{0.73+0.083\sqrt{\alpha_{}}+\alpha_{}}
    \label{vv}
\eea
where for the case of the runaway bubbles we have to substitute $\alpha\to\alpha_\infty$.
 	The factor $F_{gw,0} = \Omega_{\gamma, 0}\big( \frac{g_{s0}}{g_{s\star}}\big)^{4/3}\frac{g_\star}{g_0}= 3.57\times 10^{-5}\big(\frac{100}{g_\star}\big)^{1/3}$ converts the signal emitted at the percolation temperature to the signal we would observe today. $H_ {\text{reh}} $ and $R_\star$ are the Hubble constant and the size of the bubble at the collision (with reheating temperature correction for the Hubble constant) and the spectral shape $C(s)$ is a function determined numerically 
 	\begin{equation}
 	C(s) = s^3 \bigg(\frac{7}{4+3s^2}\bigg)^{7/2}
 	\end{equation}
 	with peak frequency 
 	\begin{equation}
 	f_{p,0} \approx 26\times 10^{-6}\bigg(\frac{1}{H_ {\text{reh}}  R_\star}\bigg)\bigg(\frac{z_p}{10}\bigg)\bigg(\frac{T_{reh}}{100 \text{ GeV}}\bigg)\bigg(\frac{g_\star}{100}\bigg)^{1/6} \text{ Hz},
 	\end{equation}
 	$g_\star$ indicates the number of relativistic degrees of freedom. 
 	Numerical simulations give $z_p \approx 10$ and $\tilde{\Omega}_{gw,0} \approx 10^{-2}$.  
\end{itemize}

\begin{figure}
	 \centering
  \includegraphics[scale=1]{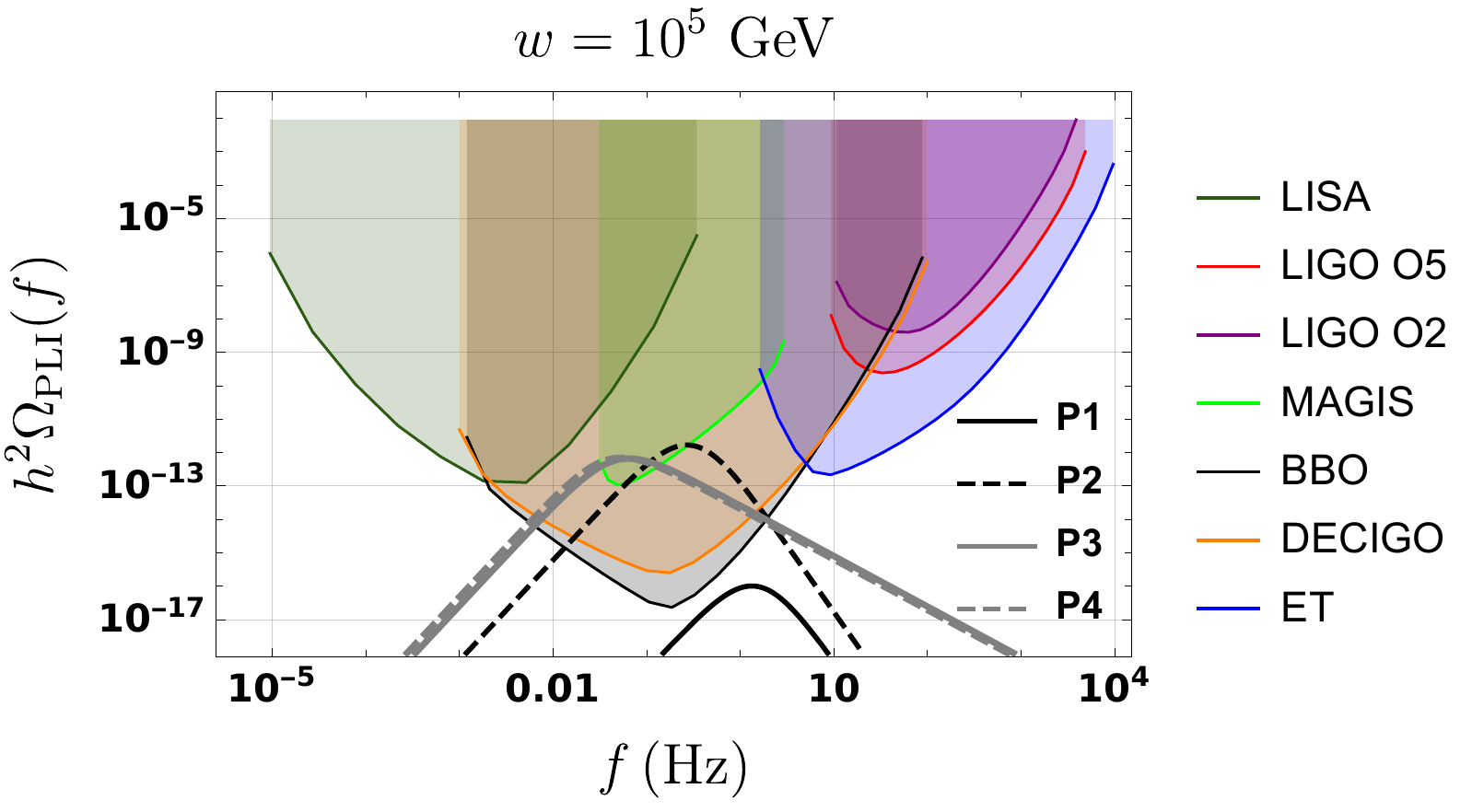}
     \caption{We represent the expected stochastic gravitational signal of the {\bf P1-P4} reference points. The spectrum is dominated by the bubble collision term for {\bf P3-P4} and by the sound waves term for {\bf P1-P2}. On the top of it, we put the PLI sensitivity curves of the coming experiments: LISA, LIGO O2 and O5, MAGIS, BBO, DECIGO and ET.
 We see that the points {\bf P2-P4} are well into the range of detection of DECIGO and BBO. {{The dominant contribution for those points is the bubble collision, or scalar field, while the efficiency factor of the sound waves contribution is largely suppressed by a factor $\sim  \frac{\alpha_{\infty}}{\alpha}$ }}. For the point {\bf P2}, 
only the sound waves component contributes to the signal with  $\kappa_{\text{fluid}} = 1$ and  we expect  {\bf P1} to be outside of the detection window as it is suppressed by $\alpha \sim 0.042$. We can also see the difference of behaviour between sound waves fuelled GW, fading as $\Omega_{sw} \sim f^{-4}$, much faster than the bubble component, fading as $\Omega_{\phi} \sim f^{-3/2}$.
     {{The signal-to-noise ratio and the sensitivity curves can be build following the recommendations of \cite{ Moore:2014lga,Aasi:2013wya,TheLIGOScientific:2014jea,
Cornish:2018dyw,Graham:2017pmn,Yagi:2011yu,Yagi:2013du,Sathyaprakash:2012jk}}
   \label{GW1}}}
\end{figure}

Armed with these expressions we can calculate the signals for the five reference points. 
The results are shown on the Fig.\ref{GW1}, where we have plotted the signals from the reference points {\bf P1-P4} on top of the power low integrated (PLI) sensitivities of the various experiments \footnote{We thank F. Sgarlata for providing the plot with experimental sensitivities.},\footnote{For alternatives for PLI see for example \cite{Alanne:2019bsm}. }.
Signal is dominated  by the sound wave contributions for the points  {\bf  P1-P2} and by the bubble collision for {\bf P3-P4}. We can see that the points with the smaller values of the coupling constants  lead to the stronger signals. This is expected since  smaller values of the couplings (which induce shallower  potentials) lead to  larger amount of  supercooling , i.e. the percolation and nucleation temperature are much lower than the critical temperature. In this case due to the larger potential energy differences between the false and true vacuum the energy release will be larger, which one clearly sees in the Table \ref{ref-values-run2} and Fig. \ref{fig:alphabeta}. Note also that the typical bubble size at the collision $\sim\frac{1}{\beta}$  increases for the smaller values of the couplings providing another factor enhancing the signal.

\section{Summary}


In summary, we recapitulate the main results of our study. We have studied the 
phase transitions in a toy model with perturbative walking dynamics  focusing on 
the possible cosmological signatures. As was mentioned in \cite{Benini:2019dfy}, the 
transition is first-order. We find that  the speed of the phase transition is controlled by the mass 
of the scalon/dilaton mode. This mode is generically the lightest field compared to 
the other ones receiving the mass during the PT, however since  its mass is only  one-loop  suppressed with respect to the tree-level masses, the mass splitting can be small.
However, the perturbative control of the  temperature 
corrections to the effective potential requires the  couplings of the model to be 
 smaller than the usual requirements of the zero temperature field  theory.
This condition makes the dilaton/scalon  particularly light compared to the other fields.
In this limit we find that  the PT occurs very slowly with significant 
amount of supercooling and detectable GW signals. We find also that increasing the couplings  leads generically  to the heavier  dilaton, a faster FOPT and smaller/vanishing GW signals.

We also compared our perturbative model with strongly-coupled models studied via holographic methods. Interestingly in both scenarios the calculations are reliable only for the light scalon/dilaton case  leading to very similar phenomenology, though in our case the supercooling { does not have to be as }strong  as in holographic models.

	It is not clear how the results of this study can be generalized for more compelling models from Beyond Standard Model prospective,  i.e.  strongly-coupled walking  theories without scalars. However, we believe that our analysis  clearly illustrates the very different cosmological signatures that can be observed  during the phase transition in models with walking dynamics.

\section*{Acknowledgments}
We would like to thank M. Serone, A. Urbano,  D Barducci,  F. Sgarlata and C.Iossa for discussions and comments.
This work was in part supported by the MIUR contract 2017L5W2PT.

\appendix
\section{Very brief review of the Benini-Iossa-Serone model}
\label{rev-BIS}
In this appendix, we briefly review  the Benini-Iossa-Serone (BIS) model  see for details the original paper \cite{Benini:2019dfy}.
 As was mentioned in the text the interest of the model is to provide a weakly-coupled realization of the walking dynamics and in particular 
the merger of two fixed points, reappearing along the imaginary  direction of the complex plane. The two loop beta function for the gauge coupling as well as the one loop function for the scalar quartic interactions are reported in Eq.\ref{run3}. Then the perturbative  Banks-Zaks fixed point appears  for  non-trivial zeros of the $\beta_{\lambda}$ function 
  	\begin{equation}
  	\lambda^\star  = \frac{\epsilon}{1 + x_s/50- 13\epsilon/2},~~~~  	22 - x_s - 4x_f = 75\epsilon.
\label{eq:lambda}
  	\end{equation}

\begin{figure}
  \begin{minipage}[c]{.05\linewidth}
       \includegraphics[scale=0.6]{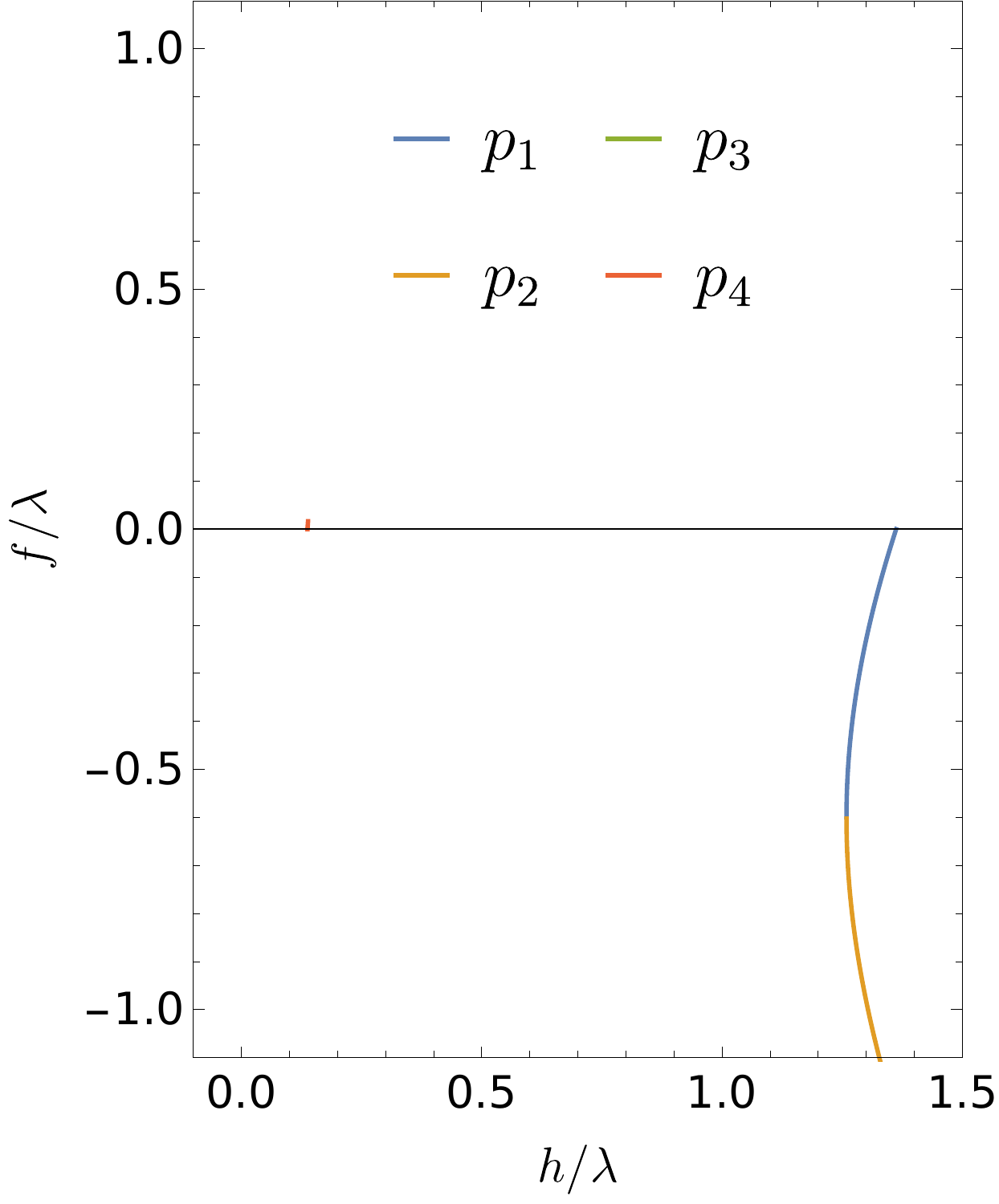}
   \end{minipage} \hfill
   \begin{minipage}[c]{.5\linewidth}
   \includegraphics[scale=0.6]{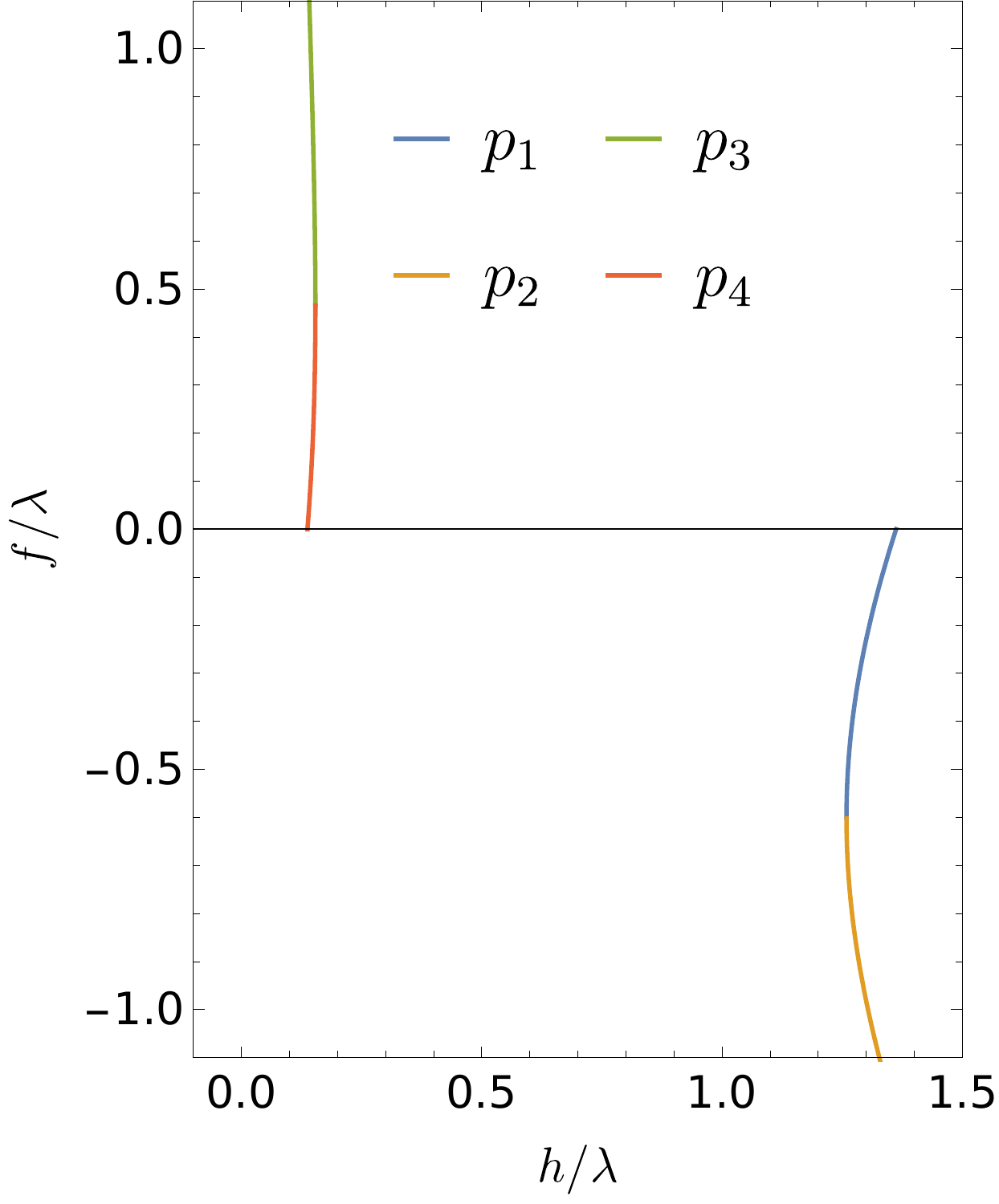}
   \end{minipage}
     \caption{Illustration of the merging of the fixed points. On the left, we give the state of the flow for $x_s \lesssim 0.08$. We see that the merging of the points $p_1$ and $p_2$ is already completed. On the right, for $x_s \lesssim 1$, the points $p_3$ and $p_4$ also merged.  }
   \label{Flow}
\end{figure}

  	We can see that the Veneziano limit decouples the $\beta_h$ of the coupling $f$. Thus we can plug $\lambda^{\star}$ into it and solve for $h$, finding in this way two fixed points for $h$ (with absolute value again parametrized by $x_s$,
  	\begin{equation}
  	h^{\star}_{\pm} = \lambda^{\star}\frac{3\pm \sqrt{6-3x_s}}{4(1+x_s)}.
  	\end{equation}
  	Again, $x_s$ parametrizes a family of fixed points that are real if $0 \leq x_s \leq  2$. Finally, plugging those values into $\beta_f$, we obtain four fixed points
  	\begin{equation}
f^{\star}_{\pm, +} = \lambda^\star (-B\pm A_+) \qquad f^\star_{\pm, -} = \lambda^\star (+B\pm A_-),
  	\end{equation}
  	where
  	\begin{equation}
  	B = \frac{\sqrt{6-3x_s}}{4},\qquad A_\pm = \frac{\sqrt{3}\sqrt{2- (13\pm 6\sqrt{6-3x_s})x_s x_s^2 -2 x_s^3}}{4(1+x_s)}. 
  	\end{equation}
  	Now we can see that $A_+$ and $A_-$ becomes complex respectively for $x_s > 0.07309 $ and $x_s < 0.8403 $. Let us thus label the four fixed points by $p_i$, 
  	\begin{equation}
  	p_1 = [\lambda^\star,h^\star_+,f^{\star}_{++}], \qquad p_3 = [\lambda^\star,h^\star_-,f^{\star}_{+-}]
  	\end{equation}
  	\begin{equation}
  	p_2 = [\lambda^\star,h^\star_+,f^{\star}_{-+}],\qquad p_4 = [\lambda^\star,h^\star_-,f^{\star}_{--}].
  	\end{equation}
  		\begin{figure}
  		\centering
 \includegraphics[scale=0.75]{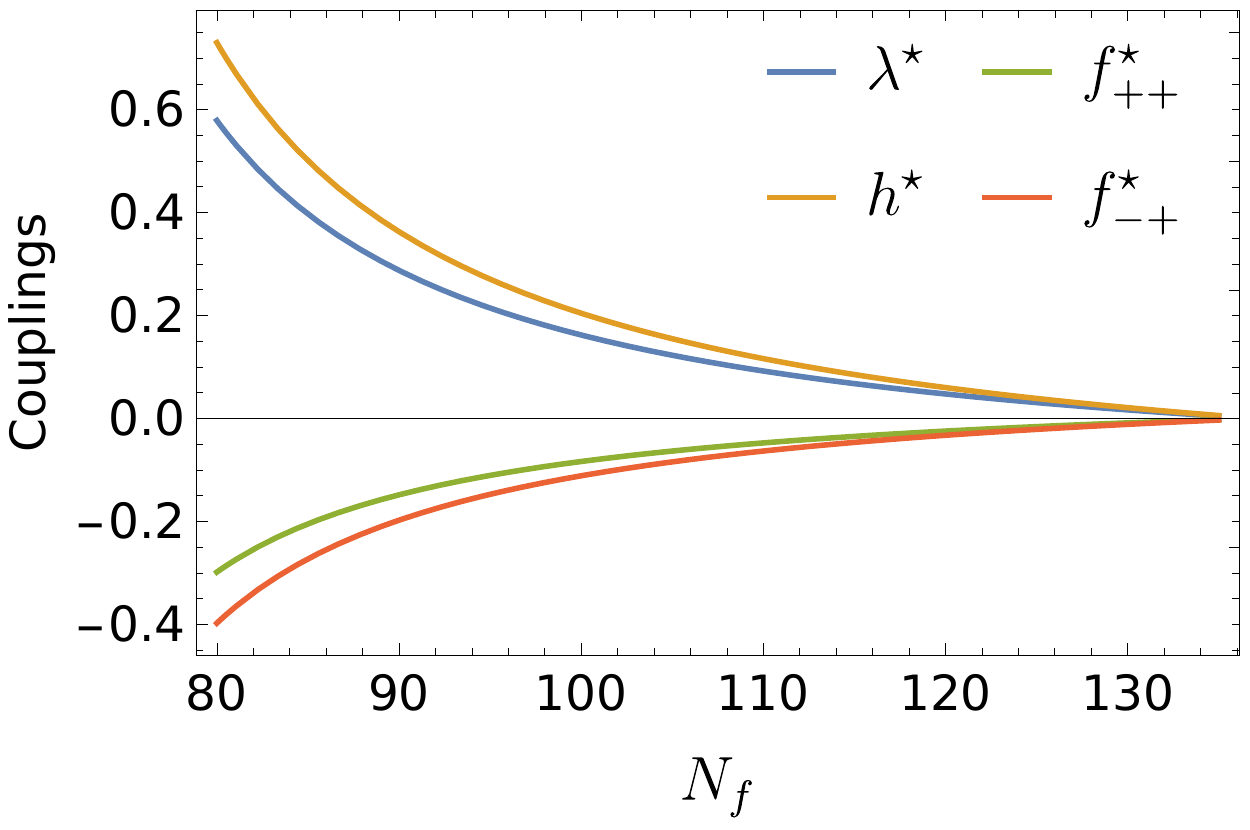}
     \caption{Values of the couplings at the fixed points $p_1 = [\lambda^\star,h^\star_+,f^{\star}_{++}]$ and $p_2 = [\lambda^\star,h^\star_+,f^{\star}_{-+}]$ for $N_s = 2$ and $N_c = 28$ fixed. As we increase the number of fermions, the absolute values of the couplings at the fixed decrease rather fast. We of course expect that this behaviour remains in the exact case. }
   \label{FPcouplings}
\end{figure} 
On the Fig.\ref{FPcouplings} we report the values of the
couplings at the fixed point for $N_s = 2, N_c = 28$.
  	On the Fig.\ref{Flow}, we can see that indeed $p_1$ and $p_2$ merge before $x_s \sim 0.08$, while the merging of $p_3$ and $p_4$ is completed much later, after $x_s \sim 1$. 
For our purpose, we will thus be interested in the first merging, for values of $x_s$ around $0.07$ At this point, we can note that this result is rather consistent with the result of \cite{Appelquist:1996dq} which estimates the exit of the conformal window around $x_f \sim 4$. 
Once two of the fixed points for the $\beta_f$ become complex we can see that the evolution of the coupling indeed has a walking behaviour see Figure \ref{fig:walking}.

Generically the  space of the BIS model  can be parametrized with three quantities, the number of colors $N_c$, the number of fermions $N_f$ and the number of scalars $N_s$.  Requiring the theory to be UV free and to pass near the complex fixed point fixes the ratio of scales separated by the walking behavior as well as the couplings constants at the instance of the spontaneous symmetry breaking. For our analysis we decided to choose the minimal number of fields where the walking phenomena is observed,  that is to say $N_s=2,N_c=25$, then the number of fermions controls  the values of the couplings (See  Eq.~\ref{eq:lambda} ) during the walking and the symmetry breaking.

\section{One field bounce dominance}
\label{sec:1dbounce}

In general, when computing the rate of tunnelling from an unstable ground state to a new ground state, we are searching for the path of least resistance from the false vacuum, where the vacuum expectation (VEV) of the higgs-like field (in our case, the scalon) is zero, to the true vacuum, where the VEV is non vanishing. It is thus a extremization problem and the task is to compute the path extremizing the Euclidean action functional 
  \begin{equation}
  S_E[\phi_i] = \int d^4x \bigg[\frac{1}{2}(\partial_\mu \phi_i)^2-V(\phi)\bigg]
  \end{equation}
  
  In the main text, we assumed that the path of least resistance from the two minima was along the scalon/dilaton direction of the potential. However, in a theory with many degrees of freedom (as the one we are considering), it does not need to be the case, as the path of least resistance could also deviate along the perpendicular directions, and "take a faster route". 

However for the model under consideration it turns out that the tunnelling only along the scalon direction is the fastest route.  We can see it by considering a specific direction perpendicular to the scalon which we can parametrize in the following way:
\begin{equation}
\phi =\l(\baa{cccc}\frac{x}{\sqrt N_s}&a/\sqrt{2}&&\\
a/\sqrt{2}&\frac{x}{\sqrt N_s}&&\\
&&\frac{x}{\sqrt N_s}&...\\
\eaa\r). 
\end{equation}

	In this decomposition, we singled out one specific perpendicular direction, $a$ and we work in the space $x-a$. Let us recall that this decomposition is done in the color-flavor space. As a consequence, the field $\phi_i^{j}$ is a $N_s \times N_c$ matrix. This specific symmetric decomposition holds for the symmetric $N_s \times N_s$ upper sub-space, where $\phi_i^{j}= \phi^{i}_j$ and $i,j \leq N_s$. 
	The tree-level potential expression
	\begin{equation}
	V[\phi] = \tilde{h} \text{Tr}\phi^{\dagger}\phi\phi^{\dagger}\phi + \tilde{f} (\text{Tr}\phi^{\dagger}\phi)^2
	\end{equation}
	becomes, by inserting the decomposition above (and keeping only terms containing the field $a$),
	\begin{equation}
	V[\phi(x,a)] = \tilde{h}\bigg(6\frac{x^2a^2}{N_s}+\frac{a^4}{2}\bigg)+ \tilde{f}\bigg(2x^2a^2+a^4\bigg). 
	\end{equation}
	Now, restricting the analysis to spontaneous symmetry breaking event, where $\tilde{h} = -N_s\tilde{f}$ (where $\tilde{h}$ is positive), the potential becomes 
	\begin{equation}
	V[x,a] = \tilde{h}\bigg(\frac{4x^2a^2}{N_s}+\frac{a^4}{2}\bigg(1-\frac{2}{N_s}\bigg)\bigg). 
	\label{pot1}
	\end{equation}

Another type of direction in the field space orthogonal to the scalon field comes from the components not residing in the $N_s\times N_s$ sub-space. 
For example we can consider the component $\phi_i^{j}, i \leq N_s, \quad j> N_s$, and call it $\phi_i^{j} \equiv b$. In this case, the only fields at hand are $\phi^{j}_i \equiv b$ and  $\phi^{i,\dagger}_j \equiv b^{\dagger}$,  $\phi^{i}_j$ and  $\phi^{j,\dagger}_i$ being out of the matrix. The potential is 
	\begin{align}
	V[x, b] &=\tilde{h}\bigg(2\frac{x^2b^2}{N_s}+b^4\bigg) +  \tilde{f}\bigg(2x^2b^2+b^4\bigg)
	\\ & \to \tilde{h}b^4\bigg(1-\frac{1}{N_s}\bigg)
	\label{pot2}
	\end{align}

	We would like now to argue that the form of this tree-level potential forces the tunnelling to happen along the scalon direction only. First, we have to recall that, at tree-level, the  potential has a flat potential along the scalon direction
\begin{equation}
V[x, a = 0,b=0]_{SB} = 0. 
\end{equation}

	The positivity of the potential \eqref{pot1},\eqref{pot2} along the $a$ and $b$-direction induces that, at tree-level, the minimum of the potential landscape is along the $x$-direction. Thus, at this order, the tunnelling will follow a straight line along the scalon direction. This conclusion still holds at higher orders as long as perturbativity is verified, thanks to the loop suppression. Therefore, even if the loop-corrections lift the scalon direction, it remains the path of least-resistance. 

\section{Properties of the phase transitions for the various values of the symmetry breaking scale. }
\label{sec:3vaiousw}
\begin{figure}
\includegraphics[scale=0.65]{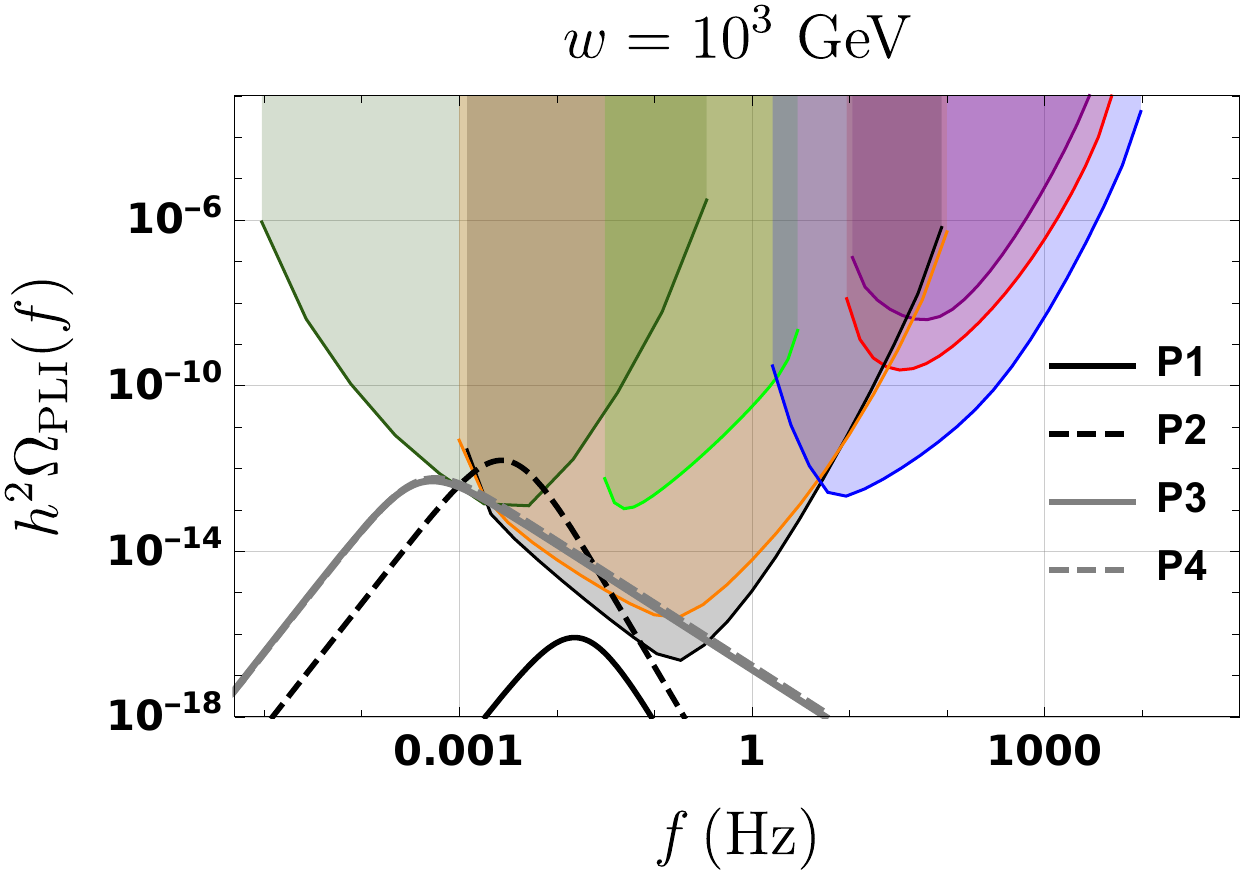}
\includegraphics[scale=0.65]{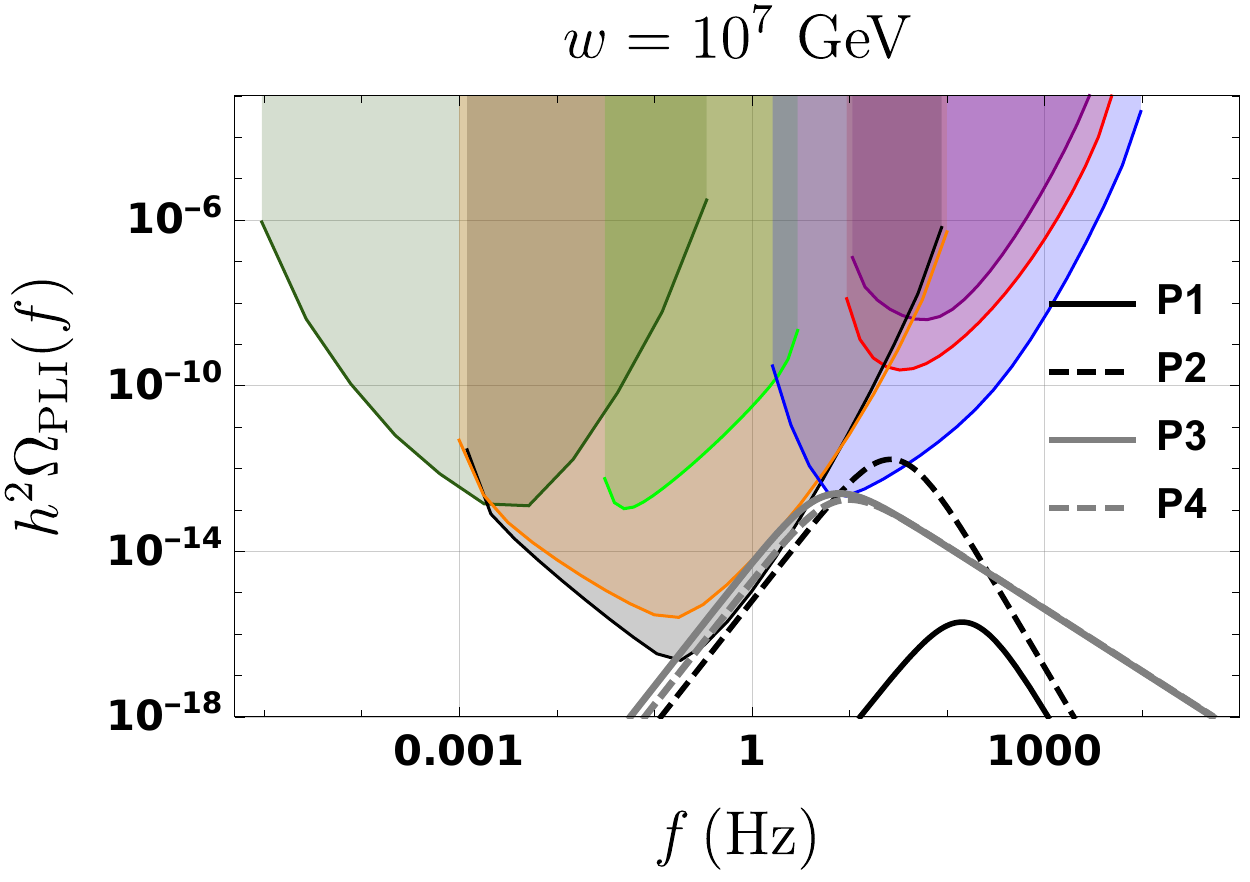}
\centering
\includegraphics[scale=0.7]{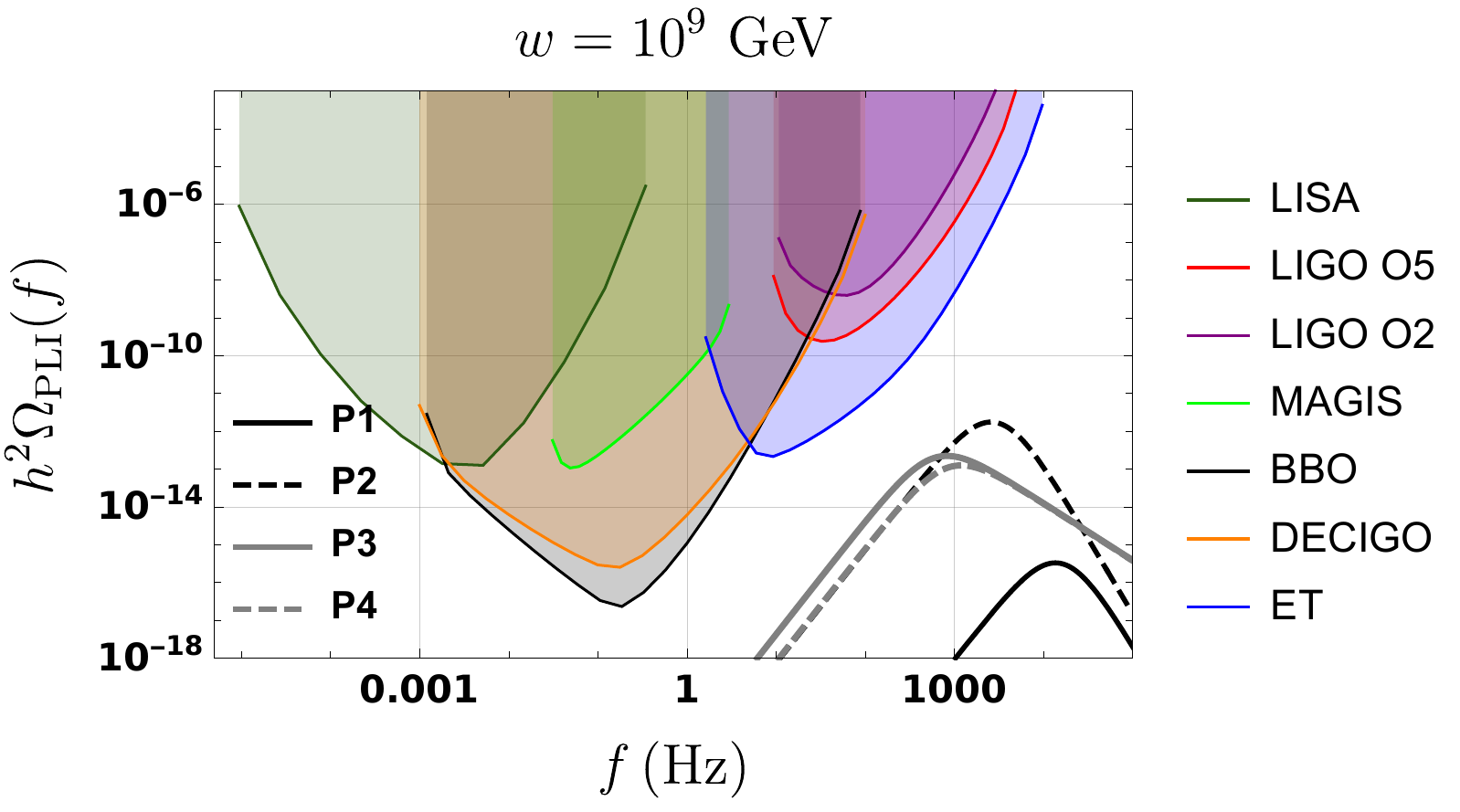}
\caption{Same as Figure \ref{GW1}   for the values of the scale $w=10^3,10^7,10^9$ GeV. \label{fig:extrapt}  }
\end{figure}

In this appendix we report the properties of the phase transition and the corresponding GW signal for the various values of the scale of the model $w$.
The results are summarized in the tables \ref{tab:extrapt} and figure \ref{fig:extrapt}.  We can see that properties of the phase transition are almost not changing with the variation of the scale $w$, so that the signal in stochastic gravitational wave background is just  shifted  towards higher or lower frequencies depending on the value of the scale $w$. Interestingly even for the value of $w=10^7$ GeV some of the experimental proposals ({\bf ET, BBO, DECIGO}) are sensitive  for the predicted signal.
\begin{table}[]
\begin{center}
\begin{tabular}{llllll}
\hline
\multicolumn{1}{|l|}{Ref. point} & \multicolumn{1}{l|}{\bf P1} & \multicolumn{1}{l|}{\bf P2} & \multicolumn{1}{l|}{\bf  P3} & \multicolumn{1}{l|}{\bf P4}  & \multicolumn{1}{l|}{\bf P5}    \\ \hline
\multicolumn{1}{|l|}{$N_f$} & \multicolumn{1}{l|}{120} & \multicolumn{1}{l|}{130}& \multicolumn{1}{l|}{133}  & \multicolumn{1}{l|}{134} & \multicolumn{1}{l|}{136} \\
\hline
\multicolumn{1}{|l|}{$\epsilon$} & \multicolumn{1}{l|}{0.0362} & \multicolumn{1}{l|}{0.0149} & \multicolumn{1}{l|}{0.0064} & \multicolumn{1}{l|}{0.00426} & \multicolumn{1}{l|}{0.00213}  \\ \hline
\multicolumn{6}{|l|}{ ~~~~~~~~~~~~~~~~~~~~~~~~~~$N_c=25$,~~~~~~~~~~~~~~~~~$N_s=2$}                                                                                               \\
\hline
\multicolumn{1}{|l|}{$\lambda$ at SB} & \multicolumn{1}{l|}{0.0473} & \multicolumn{1}{l|}{0.0166} & \multicolumn{1}{l|}{0.009} & \multicolumn{1}{l|}{0.0067} & \multicolumn{1}{l|}{0.0021} \\ \hline
\multicolumn{1}{|l|}{naive loop expansion} & \multicolumn{1}{l|}{$\sim$1.2} & \multicolumn{1}{l|}{$\sim $0.75} & \multicolumn{1}{l|}{$\sim $0.5} & \multicolumn{1}{l|}{
$\sim 0.45 $} & \multicolumn{1}{l|}{$\sim$0.3} \\ \hline
\multicolumn{1}{|l|}{$h$ at SB} & \multicolumn{1}{l|}{0.066} & \multicolumn{1}{l|}{0.023} & \multicolumn{1}{l|}{0.0126}& \multicolumn{1}{l|}{0.0093} & \multicolumn{1}{l|}{0.003} \\ \hline
\multicolumn{1}{|l|}{$f$ at SB} & \multicolumn{1}{l|}{-0.066} & \multicolumn{1}{l|}{-0.023} & \multicolumn{1}{l|}{-0.0126}& \multicolumn{1}{l|}{-0.0093} & \multicolumn{1}{l|}{-0.003} \\ \hline
%
%
\multicolumn{1}{|l|}{$T_{\text{min pert}}/w$ } & \multicolumn{1}{l|}{$- $} & \multicolumn{1}{l|}{$2\times 10^{-4} $}& \multicolumn{1}{l|}{$<10^{-10}$} & \multicolumn{1}{l|}{$<10^{-10}$} & \multicolumn{1}{l|}{$<10^{-10}$}\\ \hline
\multicolumn{1}{|l|}{$T_{\text{cr}}/w$ } & \multicolumn{1}{l|}{0.3} & \multicolumn{1}{l|}{0.167}& \multicolumn{1}{l|}{0.116} & \multicolumn{1}{l|}{0.096} & \multicolumn{1}{l|}{0.052}  \\ \hline
\multicolumn{1}{|l|}{$\frac{m^2_{\text{scalon}}}{m^2_{\text{gauge}}}$ at SB} & \multicolumn{1}{l|}{0.049} & \multicolumn{1}{l|}{0.017}& \multicolumn{1}{l|}{0.009} & \multicolumn{1}{l|}{0.0069} & \multicolumn{1}{l|}{0.0022}  \\ \hline
\multicolumn{6}{|l|}{ } \\
\multicolumn{6}{|l|}{ ~~~~~~~~~~~~~~~~~Phase transition parameters for $w=10^3$ GeV}                                                                                               \\
\multicolumn{6}{|l|}{\hspace{6 cm}{\bf \Large $\Downarrow$}}                                                                                               \\
\hline
\multicolumn{1}{|l|}{$T^{\text{nuc}}$ } & \multicolumn{1}{l|}{0.15} & \multicolumn{1}{l|}{0.0073}& \multicolumn{1}{l|}{{2.2 $\times 10^{-4}$}}  & \multicolumn{1}{l|}{$1.25\times 10^{-5}$} & \multicolumn{1}{l|}{$-$} \\ \hline
\multicolumn{1}{|l|}{$T^{\text{per}}$ } & \multicolumn{1}{l|}{0.144} & \multicolumn{1}{l|}{0.0069}& \multicolumn{1}{l|}{ {2.1 $\times 10^{-4}$}}  & \multicolumn{1}{l|}{$1.18\times 10^{-5}$} & \multicolumn{1}{l|}{$-$} \\ \hline
\multicolumn{1}{|l|}{$\alpha$ } & \multicolumn{1}{l|}{0.042} & \multicolumn{1}{l|}{730}& \multicolumn{1}{l|}{$0.8\times 10^{8}$} & \multicolumn{1}{l|}{$1.5\times 10^{13}$} & \multicolumn{1}{l|}{$-$}  \\ \hline
\multicolumn{1}{|l|}{$\beta/H=T\frac{d}{dT}\l(\frac{S_3}{T}\r)$ } & \multicolumn{1}{l|}{524} & \multicolumn{1}{l|}{382} & \multicolumn{1}{l|}{{380}}& \multicolumn{1}{l|}{{375}} & \multicolumn{1}{l|}{$-$}\\ \hline
\multicolumn{1}{|l|}{$\alpha_\infty$ } & \multicolumn{1}{l|}{0.06} & \multicolumn{1}{l|}{8}& \multicolumn{1}{l|}{$5000$} & \multicolumn{1}{l|}{$1.1\times 10^{6}$} & \multicolumn{1}{l|}{$-$}  \\ \hline

\multicolumn{6}{|l|}{ } \\
\multicolumn{6}{|l|}{ ~~~~~~~~~~~~~~~~~Phase transition parameters for $w=10^7$ GeV}                                                                                               \\
\multicolumn{6}{|l|}{\hspace{6 cm}{\bf \Large $\Downarrow$}}                                                                                               \\
\hline
\multicolumn{1}{|l|}{$T^{\text{nuc}}$ } & \multicolumn{1}{l|}{0.14} & \multicolumn{1}{l|}{0.0066}& \multicolumn{1}{l|}{{2 $\times 10^{-4}$}}  & \multicolumn{1}{l|}{$1\times 10^{-5}$} & \multicolumn{1}{l|}{$-$} \\ \hline
\multicolumn{1}{|l|}{$T^{\text{per}}$ } & \multicolumn{1}{l|}{0.134} & \multicolumn{1}{l|}{0.0062}& \multicolumn{1}{l|}{ { $1.9\times 10^{-4}$}}  & \multicolumn{1}{l|}{$0.98\times 10^{-5}$} & \multicolumn{1}{l|}{$-$} \\ \hline
\multicolumn{1}{|l|}{$\alpha$ } & \multicolumn{1}{l|}{0.057} & \multicolumn{1}{l|}{1090}& \multicolumn{1}{l|}{$1.6\times 10^{8}$} & \multicolumn{1}{l|}{$3\times 10^{13}$} & \multicolumn{1}{l|}{$-$}  \\ \hline
\multicolumn{1}{|l|}{$\beta/H=T\frac{d}{dT}\l(\frac{S_3}{T}\r)$ } & \multicolumn{1}{l|}{445} & \multicolumn{1}{l|}{370} & \multicolumn{1}{l|}{{560}}& \multicolumn{1}{l|}{{670}} & \multicolumn{1}{l|}{$-$}\\ \hline
\multicolumn{1}{|l|}{$\alpha_\infty$ } & \multicolumn{1}{l|}{0.07} & \multicolumn{1}{l|}{10}& \multicolumn{1}{l|}{$5800$} & \multicolumn{1}{l|}{$1.6\times 10^{6}$} & \multicolumn{1}{l|}{$-$}  \\ \hline

\multicolumn{6}{|l|}{ } \\
\multicolumn{6}{|l|}{ ~~~~~~~~~~~~~~~~~Phase transition parameters for $w=10^9$ GeV}                                                                                               \\
\multicolumn{6}{|l|}{\hspace{6 cm}{\bf \Large $\Downarrow$}}                                                                                               \\
\hline
\multicolumn{1}{|l|}{$T^{\text{nuc}}$ } & \multicolumn{1}{l|}{0.134} & \multicolumn{1}{l|}{0.0062}& \multicolumn{1}{l|}{{1.9 $\times 10^{-4}$}}  & \multicolumn{1}{l|}{$0.9\times 10^{-5}$} & \multicolumn{1}{l|}{$-$} \\ \hline
\multicolumn{1}{|l|}{$T^{\text{per}}$ } & \multicolumn{1}{l|}{0.128} & \multicolumn{1}{l|}{0.006}& \multicolumn{1}{l|}{ {1.84 $\times 10^{-4}$}}  & \multicolumn{1}{l|}{$0.88\times 10^{-5}$} & \multicolumn{1}{l|}{$-$} \\ \hline
\multicolumn{1}{|l|}{$\alpha$ } & \multicolumn{1}{l|}{0.069} & \multicolumn{1}{l|}{1340}& \multicolumn{1}{l|}{$2\times 10^{8}$} & \multicolumn{1}{l|}{$4.8\times 10^{13}$} & \multicolumn{1}{l|}{$-$}  \\ \hline
\multicolumn{1}{|l|}{$\beta/H=T\frac{d}{dT}\l(\frac{S_3}{T}\r)$ } & \multicolumn{1}{l|}{407} & \multicolumn{1}{l|}{360} & \multicolumn{1}{l|}{{600}}& \multicolumn{1}{l|}{{700}} & \multicolumn{1}{l|}{$-$}\\ \hline
\multicolumn{1}{|l|}{$\alpha_\infty$ } & \multicolumn{1}{l|}{0.073} & \multicolumn{1}{l|}{11}& \multicolumn{1}{l|}{$6100$} & \multicolumn{1}{l|}{$2\times 10^{6}$} & \multicolumn{1}{l|}{$-$}  \\ \hline
\end{tabular}
\end{center}
\caption{\label{tab:extrapt}
Same as Table \ref{ref-values-run2} for the values of the scale $w=10^3,10^7,10^9$ GeV.
}
\end{table}
\newpage

\bibliographystyle{JHEP}
{\footnotesize
\bibliography{biblio}}

\providecommand{\href}[2]{#2}\begingroup\raggedright\begin{thebibliography}{10}

\bibitem{Hill:2002ap}
C.~T. Hill and E.~H. Simmons {\em Phys. Rept.} {\bf 381} (2003) 235--402,
  [\href{http://arxiv.org/abs/hep-ph/0203079}{{\tt hep-ph/0203079}}]. [Erratum:
  Phys. Rept.390,553(2004)].

\bibitem{Contino:2010rs}
R.~Contino, {\it {The Higgs as a Composite Nambu-Goldstone Boson}},  in {\em
  {Physics of the large and the small, TASI 09, proceedings of the Theoretical
  Advanced Study Institute in Elementary Particle Physics, Boulder, Colorado,
  USA, 1-26 June 2009}}, pp.~235--306, 2011.
\newblock \href{http://arxiv.org/abs/1005.4269}{{\tt arXiv:1005.4269}}.

\bibitem{Bellazzini:2014yua}
B.~Bellazzini, C.~Csaki, and J.~Serra {\em Eur. Phys. J.} {\bf C74} (2014),
  no.~5 2766, [\href{http://arxiv.org/abs/1401.2457}{{\tt arXiv:1401.2457}}].

\bibitem{Holdom:1981rm}
B.~Holdom {\em Phys. Rev.} {\bf D24} (1981) 1441.

\bibitem{Yamawaki:1985zg}
K.~Yamawaki, M.~Bando, and K.-i. Matumoto {\em Phys. Rev. Lett.} {\bf 56}
  (1986) 1335.

\bibitem{Appelquist:1986an}
T.~W. Appelquist, D.~Karabali, and L.~C.~R. Wijewardhana {\em Phys. Rev. Lett.}
  {\bf 57} (1986) 957.

\bibitem{Luty:2004ye}
M.~A. Luty and T.~Okui {\em JHEP} {\bf 09} (2006) 070,
  [\href{http://arxiv.org/abs/hep-ph/0409274}{{\tt hep-ph/0409274}}].

\bibitem{Rattazzi:2008pe}
R.~Rattazzi, V.~S. Rychkov, E.~Tonni, and A.~Vichi {\em JHEP} {\bf 12} (2008)
  031, [\href{http://arxiv.org/abs/0807.0004}{{\tt arXiv:0807.0004}}].

\bibitem{Kaplan:2009kr}
D.~B. Kaplan, J.-W. Lee, D.~T. Son, and M.~A. Stephanov {\em Phys. Rev.} {\bf
  D80} (2009) 125005, [\href{http://arxiv.org/abs/0905.4752}{{\tt
  arXiv:0905.4752}}].

\bibitem{Gorbenko:2018ncu}
V.~Gorbenko, S.~Rychkov, and B.~Zan {\em JHEP} {\bf 10} (2018) 108,
  [\href{http://arxiv.org/abs/1807.11512}{{\tt arXiv:1807.11512}}].

\bibitem{Appelquist:1996dq}
T.~Appelquist, J.~Terning, and L.~C.~R. Wijewardhana {\em Phys. Rev. Lett.}
  {\bf 77} (1996) 1214--1217, [\href{http://arxiv.org/abs/hep-ph/9602385}{{\tt
  hep-ph/9602385}}].

\bibitem{Aoki:2013xza}
{\bf LatKMI} Collaboration, Y.~Aoki, T.~Aoyama, M.~Kurachi, T.~Maskawa, K.-i.
  Nagai, H.~Ohki, A.~Shibata, K.~Yamawaki, and T.~Yamazaki {\em Phys. Rev.}
  {\bf D87} (2013), no.~9 094511, [\href{http://arxiv.org/abs/1302.6859}{{\tt
  arXiv:1302.6859}}].

\bibitem{Appelquist:2014zsa}
{\bf LSD} Collaboration, T.~Appelquist et~al. {\em Phys. Rev.} {\bf D90}
  (2014), no.~11 114502, [\href{http://arxiv.org/abs/1405.4752}{{\tt
  arXiv:1405.4752}}].

\bibitem{Hasenfratz:2014rna}
A.~Hasenfratz, D.~Schaich, and A.~Veernala {\em JHEP} {\bf 06} (2015) 143,
  [\href{http://arxiv.org/abs/1410.5886}{{\tt arXiv:1410.5886}}].

\bibitem{Appelquist:2016viq}
T.~Appelquist et~al. {\em Phys. Rev.} {\bf D93} (2016), no.~11 114514,
  [\href{http://arxiv.org/abs/1601.04027}{{\tt arXiv:1601.04027}}].

\bibitem{Appelquist:2018yqe}
{\bf Lattice Strong Dynamics} Collaboration, T.~Appelquist et~al. {\em Phys.
  Rev.} {\bf D99} (2019), no.~1 014509,
  [\href{http://arxiv.org/abs/1807.08411}{{\tt arXiv:1807.08411}}].

\bibitem{Creminelli:2001th}
P.~Creminelli, A.~Nicolis, and R.~Rattazzi {\em JHEP} {\bf 03} (2002) 051,
  [\href{http://arxiv.org/abs/hep-th/0107141}{{\tt hep-th/0107141}}].

\bibitem{Bruggisser:2018mus}
S.~Bruggisser, B.~Von~Harling, O.~Matsedonskyi, and G.~Servant {\em Phys. Rev.
  Lett.} {\bf 121} (2018), no.~13 131801,
  [\href{http://arxiv.org/abs/1803.08546}{{\tt arXiv:1803.08546}}].

\bibitem{Bruggisser:2018mrt}
S.~Bruggisser, B.~Von~Harling, O.~Matsedonskyi, and G.~Servant {\em JHEP} {\bf
  12} (2018) 099, [\href{http://arxiv.org/abs/1804.07314}{{\tt
  arXiv:1804.07314}}].

\bibitem{Baratella:2018pxi}
P.~Baratella, A.~Pomarol, and F.~Rompineve {\em JHEP} {\bf 03} (2019) 100,
  [\href{http://arxiv.org/abs/1812.06996}{{\tt arXiv:1812.06996}}].

\bibitem{Agashe:2019lhy}
K.~Agashe, P.~Du, M.~Ekhterachian, S.~Kumar, and R.~Sundrum
  \href{http://arxiv.org/abs/1910.06238}{{\tt arXiv:1910.06238}}.

\bibitem{DelleRose:2019pgi}
L.~Delle~Rose, G.~Panico, M.~Redi, and A.~Tesi {\em JHEP} {\bf 04} (2020) 025,
  [\href{http://arxiv.org/abs/1912.06139}{{\tt arXiv:1912.06139}}].

\bibitem{Benini:2019dfy}
F.~Benini, C.~Iossa, and M.~Serone {\em Phys. Rev. Lett.} {\bf 124} (2020),
  no.~5 051602, [\href{http://arxiv.org/abs/1908.04325}{{\tt
  arXiv:1908.04325}}].

\bibitem{Randall:2006py}
L.~Randall and G.~Servant {\em JHEP} {\bf 05} (2007) 054,
  [\href{http://arxiv.org/abs/hep-ph/0607158}{{\tt hep-ph/0607158}}].

\bibitem{Bunk:2017fic}
D.~Bunk, J.~Hubisz, and B.~Jain {\em Eur. Phys. J.} {\bf C78} (2018), no.~1 78,
  [\href{http://arxiv.org/abs/1705.00001}{{\tt arXiv:1705.00001}}].

\bibitem{vonHarling:2017yew}
B.~von Harling and G.~Servant {\em JHEP} {\bf 01} (2018) 159,
  [\href{http://arxiv.org/abs/1711.11554}{{\tt arXiv:1711.11554}}].

\bibitem{Konstandin:2011dr}
T.~Konstandin and G.~Servant {\em JCAP} {\bf 1112} (2011) 009,
  [\href{http://arxiv.org/abs/1104.4791}{{\tt arXiv:1104.4791}}].

\bibitem{Antipin:2012kc}
O.~Antipin, S.~Di~Chiara, M.~Mojaza, E.~Molgaard, and F.~Sannino {\em Phys.
  Rev.} {\bf D86} (2012) 085009, [\href{http://arxiv.org/abs/1205.6157}{{\tt
  arXiv:1205.6157}}].

\bibitem{Hansen:2017pwe}
F.~F. Hansen, T.~Janowski, K.~Langable, R.~B. Mann, F.~Sannino, T.~G. Steele,
  and Z.-W. Wang {\em Phys. Rev.} {\bf D97} (2018), no.~6 065014,
  [\href{http://arxiv.org/abs/1706.06402}{{\tt arXiv:1706.06402}}].

\bibitem{Banks:1981nn}
T.~Banks and A.~Zaks {\em Nucl. Phys.} {\bf B196} (1982) 189--204.

\bibitem{Caswell:1974gg}
W.~E. Caswell {\em Phys. Rev. Lett.} {\bf 33} (1974) 244.

\bibitem{Miransky:1984ef}
V.~A. Miransky {\em Nuovo Cim.} {\bf A90} (1985) 149--170.

\bibitem{Weinberg:1973am}
E.~J. Weinberg, {\em {Radiative corrections as the origin of spontaneous
  symmetry breaking}}.
\newblock PhD thesis, Harvard U., 1973.
\newblock \href{http://arxiv.org/abs/hep-th/0507214}{{\tt hep-th/0507214}}.

\bibitem{Randall:1999ee}
L.~Randall and R.~Sundrum {\em Phys. Rev. Lett.} {\bf 83} (1999) 3370--3373,
  [\href{http://arxiv.org/abs/hep-ph/9905221}{{\tt hep-ph/9905221}}].

\bibitem{Goldberger:1999uk}
W.~D. Goldberger and M.~B. Wise {\em Phys. Rev. Lett.} {\bf 83} (1999)
  4922--4925, [\href{http://arxiv.org/abs/hep-ph/9907447}{{\tt
  hep-ph/9907447}}].

\bibitem{Curtin:2016urg}
D.~Curtin, P.~Meade, and H.~Ramani {\em Eur. Phys. J.} {\bf C78} (2018), no.~9
  787, [\href{http://arxiv.org/abs/1612.00466}{{\tt arXiv:1612.00466}}].

\bibitem{Comelli:1996vm}
D.~Comelli and J.~R. Espinosa {\em Phys. Rev.} {\bf D55} (1997) 6253--6263,
  [\href{http://arxiv.org/abs/hep-ph/9606438}{{\tt hep-ph/9606438}}].

\bibitem{Weinberg:1974hy}
S.~Weinberg {\em Phys. Rev.} {\bf D9} (1974) 3357--3378.

\bibitem{Arnold:1992rz}
P.~B. Arnold and O.~Espinosa {\em Phys. Rev.} {\bf D47} (1993) 3546,
  [\href{http://arxiv.org/abs/hep-ph/9212235}{{\tt hep-ph/9212235}}]. [Erratum:
  Phys. Rev.D50,6662(1994)].

\bibitem{Ellis:2018mja}
J.~Ellis, M.~Lewicki, and J.~M. No \href{http://arxiv.org/abs/1809.08242}{{\tt
  arXiv:1809.08242}}. [JCAP1904,003(2019)].

\bibitem{Dine:1992wr}
M.~Dine, R.~G. Leigh, P.~Y. Huet, A.~D. Linde, and D.~A. Linde {\em Phys. Rev.}
  {\bf D46} (1992) 550--571, [\href{http://arxiv.org/abs/hep-ph/9203203}{{\tt
  hep-ph/9203203}}].

\bibitem{Bodeker:2009qy}
D.~Bodeker and G.~D. Moore {\em JCAP} {\bf 0905} (2009) 009,
  [\href{http://arxiv.org/abs/0903.4099}{{\tt arXiv:0903.4099}}].

\bibitem{Bodeker:2017cim}
D.~Bodeker and G.~D. Moore {\em JCAP} {\bf 1705} (2017), no.~05 025,
  [\href{http://arxiv.org/abs/1703.08215}{{\tt arXiv:1703.08215}}].

\bibitem{Espinosa:2010hh}
J.~R. Espinosa, T.~Konstandin, J.~M. No, and G.~Servant {\em JCAP} {\bf 1006}
  (2010) 028, [\href{http://arxiv.org/abs/1004.4187}{{\tt arXiv:1004.4187}}].

\bibitem{Caprini:2019egz}
C.~Caprini et~al. \href{http://arxiv.org/abs/1910.13125}{{\tt
  arXiv:1910.13125}}.

\bibitem{Cutting:2018tjt}
D.~Cutting, M.~Hindmarsh, and D.~J. Weir {\em Phys. Rev.} {\bf D97} (2018),
  no.~12 123513, [\href{http://arxiv.org/abs/1802.05712}{{\tt
  arXiv:1802.05712}}].

\bibitem{Alanne:2019bsm}
T.~Alanne, T.~Hugle, M.~Platscher, and K.~Schmitz
  \href{http://arxiv.org/abs/1909.11356}{{\tt arXiv:1909.11356}}.

\bibitem{Moore:2014lga}
C.~J. Moore, R.~H. Cole, and C.~P.~L. Berry {\em Class. Quant. Grav.} {\bf 32}
  (2015), no.~1 015014, [\href{http://arxiv.org/abs/1408.0740}{{\tt
  arXiv:1408.0740}}].

\bibitem{Aasi:2013wya}
{\bf KAGRA, LIGO Scientific, VIRGO} Collaboration, B.~P. Abbott et~al. {\em
  Living Rev. Rel.} {\bf 21} (2018), no.~1 3,
  [\href{http://arxiv.org/abs/1304.0670}{{\tt arXiv:1304.0670}}].

\bibitem{TheLIGOScientific:2014jea}
{\bf LIGO Scientific} Collaboration, J.~Aasi et~al. {\em Class. Quant. Grav.}
  {\bf 32} (2015) 074001, [\href{http://arxiv.org/abs/1411.4547}{{\tt
  arXiv:1411.4547}}].

\bibitem{Cornish:2018dyw}
T.~Robson, N.~J. Cornish, and C.~Liug {\em Class. Quant. Grav.} {\bf 36}
  (2019), no.~10 105011, [\href{http://arxiv.org/abs/1803.01944}{{\tt
  arXiv:1803.01944}}].

\bibitem{Graham:2017pmn}
{\bf MAGIS} Collaboration, P.~W. Graham, J.~M. Hogan, M.~A. Kasevich,
  S.~Rajendran, and R.~W. Romani \href{http://arxiv.org/abs/1711.02225}{{\tt
  arXiv:1711.02225}}.

\bibitem{Yagi:2011yu}
K.~Yagi, N.~Tanahashi, and T.~Tanaka {\em Phys. Rev.} {\bf D83} (2011) 084036,
  [\href{http://arxiv.org/abs/1101.4997}{{\tt arXiv:1101.4997}}].

\bibitem{Yagi:2013du}
K.~Yagi {\em Int. J. Mod. Phys.} {\bf D22} (2013) 1341013,
  [\href{http://arxiv.org/abs/1302.2388}{{\tt arXiv:1302.2388}}].

\bibitem{Sathyaprakash:2012jk}
B.~Sathyaprakash et~al. {\em Class. Quant. Grav.} {\bf 29} (2012) 124013,
  [\href{http://arxiv.org/abs/1206.0331}{{\tt arXiv:1206.0331}}]. [Erratum:
  Class. Quant. Grav.30,079501(2013)].

\end{thebibliography}\endgroup

\end{document}